\documentclass[journal=jctcce,manuscript=article]{achemso}
\usepackage[utf8]{inputenc}
\usepackage{ulem}
\usepackage{amsmath}
\usepackage{gensymb}
\usepackage{xcolor}
\usepackage{graphicx}
\usepackage{amsthm}
\usepackage{dcolumn}
\usepackage{epsfig}
\usepackage{bm}
\usepackage{booktabs}
 
\author{Porhouy Minh}
\affiliation{Department of Chemistry, University of Minnesota, Minnesota 55455, United States}
\author{Steven W. Hall}
\affiliation{Department of Chemical Engineering and Materials Science, University of Minnesota, Minnesota 55455, United States}
\author{Ryan S. DeFever}
\affiliation{Department of Chemical and Biomolecular Engineering, Clemson University, South Carolina 29634, United States}
\author{Sapna Sarupria}
\email{sarupria@umn.edu}
\affiliation{Department of Chemistry, University of Minnesota, Minnesota 55455, United States}

\title{Crystal Nucleation Kinetics and Mechanism: Influence of Interaction Potential}
\date{\today}

\begin{document}
\maketitle

\begin{tocentry}
  \includegraphics[width=0.5\textwidth]{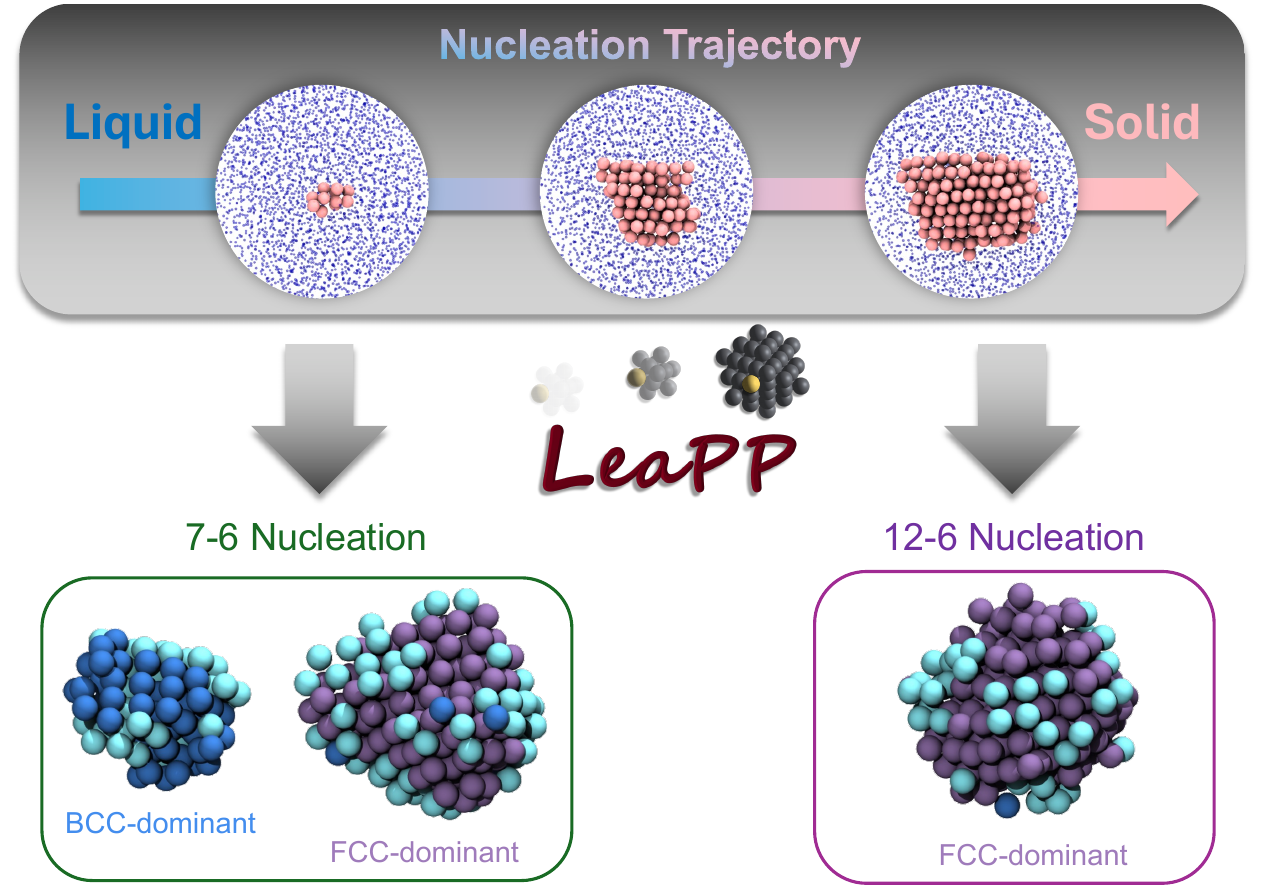}
\end{tocentry}

\begin{abstract}

Modulating liquid-to-solid transitions and the resulting crystalline structure for tailored properties is much desired. Colloidal systems are exemplary to this end, but the fundamental knowledge gaps in relating the influence of intermolecular interactions to crystallization behavior continue to hinder progress. In this study, we address this knowledge gap by studying nucleation and growth in systems with modified Lennard-Jones potential. Specifically, we study the commonly used 12--6 potential and a softer 7--6 potential. The thermodynamic state point for the study is chosen such that both systems are investigated at the same level of supercooling and pressure. Under these conditions, we find that the nucleation rate for both systems is comparable. Interestingly, the nucleation pathways and resulting crystal structures are different. In the 12-6 system, nucleation and growth occur predominantly through the FCC structure. Softening the potential alters the critical nucleus composition and introduces two distinct nucleation pathways. One pathway predominantly leads to the nucleus with a body-centered cubic (BCC) structure, while the other favors the face-centered cubic (FCC) arrangement. Our study illustrates that polymorph selection can be achieved through modifications to intermolecular interactions without impacting nucleation kinetics. The results have significant implications in designing approaches for polymorph selection and modulating self-assembly mechanisms. 
    
\end{abstract}
    
\section{Introduction}

Intermolecular interactions fundamentally affect the structural, thermodynamic, and kinetic properties of fluids and materials. In molecular simulations, these interactions are approximated by
potentials. One widely-used choice to represent van der Waals forces is the 12--6 Lennard-Jones (LJ) potential:
\begin{equation}
V(r) = 4\epsilon\Bigg[\Big(\frac{\sigma}{r}\Big)^{12} - \Big(\frac{\sigma}{r}\Big)^{6}\Bigg],
\label{eq:lj}
\end{equation}
where $V$ is the potential energy of two particles $r$ distance apart,
$\epsilon$ is the well depth, and $2^{1/6}\sigma$ is the location of
the minimum in the potential. The potential can be approximately separated into a
short-range repulsive component (at $r<2^{1/6}\sigma$) and a longer-range
attractive component (at $r>2^{1/6}\sigma$). In the van der Waals picture of dense
non-associated liquids,\cite{Chandler_Van_1983,Weeks_Role_1971}
short-range repulsions of the interaction potential determine the
structure and dynamics of molecules, while the attractive tail acts as
a mean field that determines the liquid density.

Though the LJ potential has proven widely applicable, some systems such as polymer fluids and colloidal particles may be more accurately described by ``softer'' potentials.\cite{Likos:01:PhysRep,Stillinger:76:JCP,Giovambattista:11:JPCB} In this context, ``softer'' indicates an interaction potential that exerts less force (i.e., gentler slope of $V(r)$ vs. $r$) at distances near the potential minimum. For instance, colloidal systems are often represented by effective interaction potentials\cite{Likos:01:PhysRep,Boles_Self_2016} that describe the combined effects of charge, charge screening, depletion, hard sphere repulsion, and surface functionalization by ligands. The influence of interaction potentials is particularly relevant to colloidal systems, since particle size, shape, and surface functionalization can be precisely modified to influence self-assembly.\cite{Yethiraj_Tunable_2007,Palberg_Crystallization_2014,Boles_Self_2016,Li_Assembly_2016} Therefore, tailored crystalline (self-assembled) structures of colloidal particles can be achieved by tuning the interactions between them. However, there is limited understanding of the manifestations of intermolecular interactions on the mechanisms and rates of crystal formation, which hinders our ability to leverage the tunability of intermolecular interactions for desired structures.

Simulations provide a means to tweak aspects of the potential in isolation, allowing targeted investigations of their effects. The effects of softening the LJ potential on liquid structure and dynamics and solid--liquid equilibrium have been investigated previously.\cite{Shi_Structure_2011, Ahmed_Solid_2009, Barroso2012JCP, Kofke1995MP, Ning2016AIPAdv} Shi \textit{et al.}\cite{Shi_Structure_2011} studied the effects of softness for a family of LJ potentials on the behavior of a binary glass-forming mixture of particles at constant density. Increasing softness had little effect on liquid structure but led to increasingly negative configurational potential energy due to lower potential energy in the tail. Higher diffusivities were also observed with increasing softness. Ahmed and Sadus\cite{Ahmed_Solid_2009} calculated the phase diagrams for single-component systems with a slightly different family of LJ potentials. At lower pressures, the melting temperature increased with softness, whereas the opposite trend was observed at higher pressures. Interestingly, the choice of softness was shown to influence the liquid structure at the freezing point. Upon increasing softness, the radial distribution peaks shift to smaller radial distances, thereby increasing the liquid density. Overall, these studies show that thermodynamic and dynamic properties are affected by potential softness.

Vald\`es \textit{et al.} investigated the thermodynamics and crystallization kinetics for a family of LJ potentials where only the attractive tail beyond the minimum was modified, without changing the repulsive region.\cite{Valds_Crystallization_2018} They found that stronger attractions led to higher melting temperatures and an increased driving force for nucleation for a given temperature, resulting in higher nucleation rates. At lower supercoolings, the rates appeared consistent with classical nucleation theory (CNT), while rates at higher supercoolings were overestimated by CNT.\cite{Valds_Crystallization_2018} Separately, Auer and Frenkel\cite{Frenkel:02:JPCM} examined the surface free energy and nucleation rate of a family of repulsive Yukawa potentials, where softness of interactions was tuned via the inverse charge screening length. Initially, as the softness increased, the liquid--FCC surface free energy decreased, and the nucleation rate increased until a maximum (minimum) in the rate (surface free energy) was reached. Thereafter, the trends reversed as the softness was further increased.

Mechanisms of nucleation in systems of the LJ and other isotropic potentials have also been explored.\cite{ten_Wolde_Numerical_1995,ten_Wolde_Numerical_1996,Desgranges_Controlling_2007,Trudu_Freezing_2006,Moroni_Interplay_2005,Beckham_Optimizing_2011,Travesset_Phase_2014,Desgranges_Molecular_2006,Delhommelle:07:JCP,Delhommelle:07:JPCB,Dijkstra2025JCP, Dijkstra2021ACSNano} Although the FCC solid is the thermodynamically stable phase for the LJ potential at moderate supercooling,\cite{Travesset_Phase_2014} the consensus is that the nascent nuclei are primarily BCC.\cite{ten_Wolde_Numerical_1995,Desgranges_Controlling_2007,Desgranges_Molecular_2006,Delhommelle:07:JCP,Delhommelle:07:JPCB} As nuclei grow, their core converts to an FCC structure, while their surface continues to retain some BCC character. Notably, there appears to be an interplay between the size and structure of the critical nuclei.\cite{Moroni_Interplay_2005, Beckham_Optimizing_2011} Smaller nuclei with greater FCC character were found to be equally likely to grow as larger nuclei with more BCC character.\cite{Moroni_Interplay_2005} The structural preference between FCC and BCC within a nucleus gets even more interesting when the interaction potential is softened. Desgranges and Delhommelle\cite{Delhommelle:07:JCP} showed that for the softer repulsive Yukawa potential, where BCC is the most stable phase, the critical nucleus was primarily BCC and largely remained that way through growth. For the more repulsive potential with a stable FCC phase, the critical nucleus was a mixture of FCC and BCC, becoming primarily FCC during the growth phase.

The studies above highlight that changes to the intermolecular potential can result in varying degrees of changes in the thermodynamics and kinetics of crystallization. In most of these studies, however, the changes observed in crystallization kinetics were conflated effects of both modifications to the potential and shifts in the thermodynamic driving forces. 
Furthermore, the influence of softening both repulsive and attractive components of the interaction potential has not been investigated. Understanding these aspects is important to complete the connection between intermolecular interactions and the resulting nucleation behavior. We address this knowledge gap by investigating the effect of softening both the repulsive and attractive interactions of the LJ potential on the kinetics and mechanism of crystal nucleation under the same driving force.

We study nucleation in systems with intermolecular interactions governed by a modified n-6 potential\cite{Shi_Structure_2011} (namely 12--6 and 7--6 potentials). Using a suite of methods, including replica exchange transition interface sampling (RETIS)\cite{vanErp:07:PRL, vanErp:17:JCP}, seeding\cite{Espinosa_Seeding_2016}, and classical nucleation theory\cite{VolmerWeber1926,Farkas1927,BeckerDoring1935, Zeldovich1943,Sosso_Crystal_2016,Debenedetti_Metastable_1996,Kashchiev_Nucleation_2000}, we characterize the effect of changing the n-6 potential on the nucleation kinetics. Furthermore, by employing the recently developed LeaPP methodology\cite{SarupriaJCTC2025} to characterize the nucleation pathways, we discover the interplay between intermolecular interactions and the resulting polymorph structure. Our results indicate that the different interaction potentials result in comparable nucleation rates at the same driving force but follow distinct nucleation mechanisms. Softening the potential stabilizes the body-centered cubic (BCC) structure and introduces an alternative nucleation pathway distinct from the face-centered cubic (FCC) pathway observed in the 12--6 LJ system. These insights demonstrate that the driving forces, nucleation rates, and resulting structures for crystallization and self-assembly can be controlled through a careful modulation of the intermolecular interactions.

\section{Methods}

\subsection{Model Potentials}

\begin{figure}[h!]
\begin{center}
\includegraphics[width=0.4\textwidth]{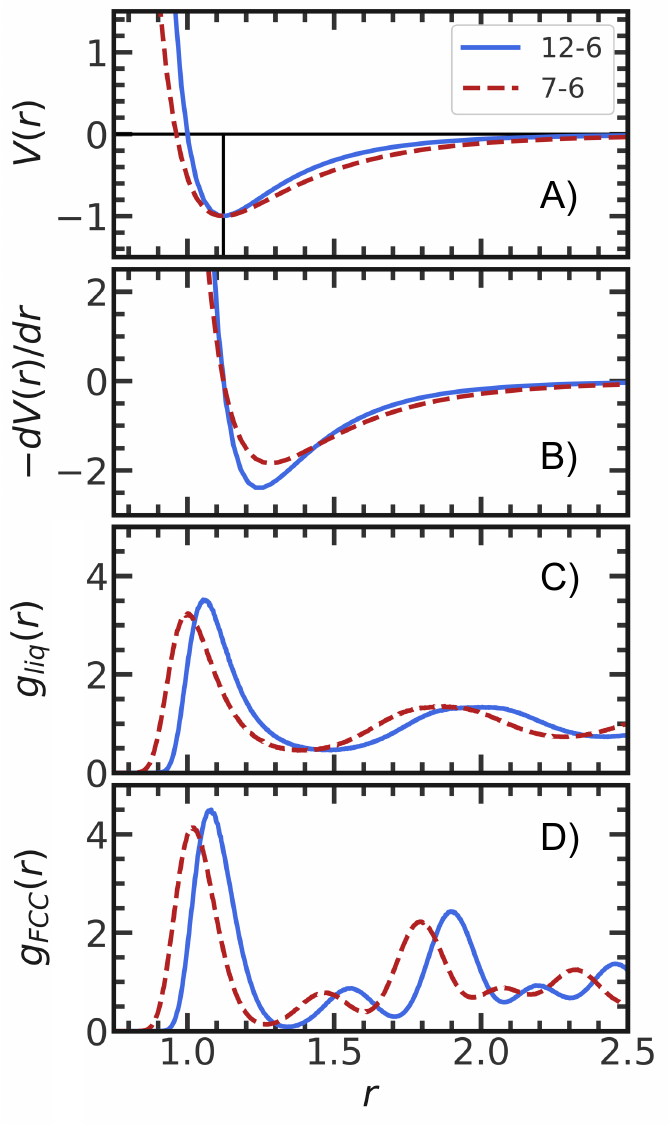}
\vspace{-0.2in}
\caption{(A) The potentials studied, (B) the negative of their derivatives, (C) liquid RDF ($g_{liq}(r)$) at $p=5$, $T=0.78T_{\text{m}}$, and (D) FCC crystal RDF ($g_{FCC}(r)$) at $p=5$, $T=0.78T_{\text{m}}$.}
\label{fig:potential}
\end{center}
\end{figure}

The general form of the modified $n$--6 potential is given by\cite{Shi_Structure_2011}
\begin{equation}
V(r)=\epsilon\Bigg[\theta\Big(\frac{\sigma}{r}\Big)^n-\psi\Big(\frac{\sigma}{r}\Big)^6\Bigg],
\end{equation}
where $n$ is the repulsive exponent, $\sigma$ is the distance at which
the potential energy is zero, and $\epsilon$ is the depth of the
potential minimum. $\theta$ and $\psi$ are chosen such that the minima of the
potentials are located at $r=2^{1/6}\sigma$. Here, we consider
potentials with $n=12$ and $n=7$ (Fig.~\ref{fig:potential}). When $n=12$, $\theta=\psi=4$ and the standard 
LJ potential is recovered. When $n=7$, $\theta=13.468$ and
$\psi=14.000$. The $n=7$ potential is softer than the LJ potential
($n=12$). Following the standard convention, all reported values are
given in reduced units: energy and free energy in units of $\epsilon$,
length in units of $\sigma$, and time in units of
$\tau=\sigma(m/\epsilon)^{1/2}$ where $m$ is the particle
mass. Temperature ($T$), pressure ($p$), density ($\rho$), surface
free energy ($\gamma$), and chemical potential ($\mu$) are expressed in units of
$\epsilon/k_{\text{B}}$, $\epsilon/\sigma^3$, $\sigma^{-3}$, $\epsilon/\sigma^2$, and $\epsilon$, respectively.

Plots of the two potentials, their derivatives (i.e., force,
$F=-dV(r)/dr$), and their liquid and FCC radial distribution functions (RDFs) are shown in Fig.~\ref{fig:potential}. The 12--6 potential displays stronger close-range repulsion and stronger attraction at distances between the potential minimum and $\sim$1.5$\sigma$. Beyond this distance, the attraction of the 7--6 potential is greater (Fig.~\ref{fig:potential}B). The lower magnitude force near the potential minimum results in the `softer' potential; particles experience a weaker restorative force when they deviate from the location of their potential energy minima.

The liquid and FCC RDFs in Fig.~\ref{fig:potential}C and Fig.~\ref{fig:potential}D illustrate how the differences in the two potentials manifest in the liquid and solid structures. The peak locations of the RDFs for the 7--6 potential are shifted to closer distances for both liquid and solid, indicative of an increase in the density. The first peak height and width of the RDFs of the 7--6 potential are slightly lower and wider, suggesting that particles within the first neighbor shell are more closely packed, yet more diffusive (squishy). The same trend is observed for the second peak of the FCC RDF. Interestingly, the second peak width of the liquid RDF for the 7--6 potential is slightly narrower compared to the 12--6. This suggests increased structural order in the second neighbor shell of the liquid, possibly due to the stronger attraction of the 7--6 tail at distances beyond $\sim$1.5$\sigma$. When only modifying the attractive tail of LJ-like potentials, Vald\`{e}s \textit{et al.}\cite{Valds_Crystallization_2018} also found that the peak locations shifted to closer distances with increasing softness in the potential tail. However, the magnitude of the peak location shift was smaller, and no change to the peak heights or widths was observed.

\subsection{Simulation Details}

We conducted direct coexistence simulations\cite{Fernndez_The_2006} to identify the melting temperatures, for both potentials, over pressures ranging from 0 to $20\epsilon/\sigma^{3}$ (see Fig. S1). We then selected a pressure at which the two systems exhibited the same $T_m$ on their respective coexistence curves. The state point for all subsequent simulations was set to $p=5$ and $T=0.78T_{m}$, yielding an equivalent degree of (moderate) supercooling in both systems. This choice lets us isolate the influence of the $n$--6 modification, since both systems are investigated at an identical level of supercooling and pressure. Phase equilibrium simulation protocols are provided in the SI.

The MD simulations were performed with 8192 particles using GROMACS v5.1.2\cite{Hess_GROMACS_2008,Abraham_GROMACS_2015}. The leapfrog integrator was used with a time step of 0.001$\tau$. Periodic boundaries were applied in all three dimensions and linear center of mass motion was removed every 100 steps. The GROMACS implementation of the force-switch was applied to the potentials to smoothly bring the forces to zero from the switching distance of 3.0$\sigma$ to the cut-off distance of 3.5$\sigma$. Neighbor lists were updated every 10 steps within a distance cut-off of 4.0$\sigma$. Temperature and pressure were controlled with the Nos\'{e}--Hoover thermostat~\cite{Hoover_Canonical_1985} and Parrinello--Rahman barostat,~\cite{Parrinello_Polymorphic_1981} respectively. Coupling constants for the thermostat and barostat were $1\tau$ and $5\tau$, respectively.

\subsubsection{Replica Exchange Transition Interface Sampling}

Crystal nucleation is a rare event in that the waiting time to observe the phase transition is much longer than the timescale of the transition. As such, at moderate supercoolings such as $T=0.78T_m$, straightforward MD simulations initiated in the liquid phase will remain liquid and not transition to the solid phase on readily accessible timescales. To overcome this limitation, replica exchange transition interface sampling (RETIS)\cite{vanErp:07:PRL, vanErp:17:JCP} was employed to enhance the sampling of nucleation events with unbiased dynamics. The resulting nucleation trajectories were stored and subsequently analyzed to compute the nucleation rate and to uncover the underlying mechanism of crystal nucleation. 

RETIS was performed along an order parameter ($\lambda$) that corresponded to the size of the largest solid nucleus in a configuration. This largest nucleus size was calculated by identifying the number of crystalline particles in the largest solid nucleus, n$_\text{tf}$, via the Steinhardt $q_6$ bond order parameter (see Section S5 in the SI for details).\cite{ten_Wolde_Numerical_1996, tenWolde:05:PRL,SarupriaJCTC2025,SteinhardtPRB1983} In both the 7--6 and 12--6 systems, the transition region was divided into 17 interface ensembles. For 7--6, interfaces were positioned at $\lambda$ = \{45, 55, 70, 85, 100, 115, 130, 145, 170, 190, 220, 250, 280, 320, 360, 410, 520\}, and the final interface was placed at $\lambda = 720$.\cite{SarupriaJCTC2025} For 12--6, the interfaces were placed at $\lambda$ = \{45, 55, 70, 80, 90, 100, 110, 130, 145, 160, 180, 195, 215, 235, 270, 330, 420\}, the final interface was placed at $\lambda = 600$. The position of each interface was chosen such that there was at least 10\% overlap in the crossing histograms between neighboring interfaces (i.e., $P(\lambda_{i+1}|\lambda_{i} )\geq0.1$).\cite{SarupriaJCP2022} 

For each interface, 5000 moves were performed with 50\% shooting and 50\% exchange moves. We discarded the first 600 trajectories of each interface ensemble to remove bias caused by their initial trajectories. Then, trajectories were collected after every 10 moves to ensure sufficient trajectory decorrelation. This generated 440 decorrelated trajectories per interface ensemble for analysis. Trajectory decorrelation and sampling convergence were assessed following the best practices outlined in Ref.~\citenum{SarupriaJCP2022}. Crossing probability histograms for both systems are shown in Fig. S4 of the SI. 

\subsection{Classical Nucleation Theory}\label{sec:CNT}
While RETIS provides a direct estimate of the nucleation rate, classical nucleation theory (CNT)\cite{VolmerWeber1926,Farkas1927,BeckerDoring1935,Zeldovich1943,Sosso_Crystal_2016,Debenedetti_Metastable_1996,Kashchiev_Nucleation_2000} offers a framework to probe the underlying thermodynamic and kinetic factors. We leverage CNT in conjunction with seeding\cite{Espinosa_Seeding_2016,Zimmermann_NaCl_2018} to investigate the factors that contribute to the nucleation rate. According to CNT, the free energy required to form a solid nucleus of size $n$ in a metastable liquid is given by

\begin{equation}
\Delta G(n)=-|\Delta\mu|n+\gamma \phi n^{2/3}
\label{eq:g-cnt}
\end{equation}
where $\Delta\mu$ is the solid--liquid chemical potential difference, $\gamma$ is the solid--liquid surface tension, $\phi$ is a shape factor, and the product $\phi n^{2/3}$ is the surface area of the nucleus. For spherical nuclei, $\phi = (36\pi/\rho_{\text{s}}^2)^{1/3}$, where $\rho_{\text{s}}$ is the equilibrium density of the solid phase nucleating from the liquid. By maximizing $\Delta G$ with respect to $n$, the barrier height is given as:

\begin{equation}
\Delta G_c =\frac{n_{\text{c}} |\Delta\mu|}{2},
\label{eq:critbarrier}
\end{equation}
where $\Delta G_c$ is the free energy required to form a critical nucleus of size $n_{\text{c}}$. For spherical nuclei,

\begin{equation}
n_{\text{c}}=\frac{32\pi}{3}\frac{\gamma^3}{\rho^2_{\text{s}}|\Delta\mu|^3}.
\label{eq:critsize}
\end{equation}
The nucleation rate is given by

\begin{equation}
J=\rho_{\text{l}} Z D \exp{[-\Delta G_{\text{c}} /k_\text{B}T]},
\label{eq:seed-rate}
\end{equation}
where $Z$ is the Zeldovich correction factor and $D$ is the attachment
rate to the critical nucleus (diffusivity of the coordinate $n$).  The
Zeldovich factor can be calculated as

\begin{equation}
Z=\sqrt{|\Delta\mu|/(6\pi k_\text{B}T n_{\text{c}})}.
\label{eq:zeld}
\end{equation}

To apply Eqs. \ref{eq:g-cnt}--\ref{eq:zeld} for nucleation rate estimation, we need to estimate $\rho_s$, $\rho_l$, $| \Delta \mu |$, $\gamma$, and $D$.  $\rho_s$ and $\rho_l$ can be estimated from an MD simulation of the pure solid and liquid phase, respectively, and $| \Delta \mu |$ can be estimated from thermodynamic integration~\cite{Vega:14:JCP} using the equation:

\begin{equation}
\Delta\mu(T^{*}) = T^{*}\int_{T^{*}}^{T_{\text{m}}} \frac{\Delta H(T)}{T^2} dT,
\end{equation} where $\Delta H$ is the solid-liquid enthalpy difference at $T$ and $T^{*}$ is the supercooled temperature of interest. $\gamma$ and $D$ are more difficult to estimate. 

\subsubsection{Estimation of $\gamma$ and $D$ from seeding}\label{sec:seed-2}

The general idea of seeding is to use simulations of seeds of the crystalline phase, which are placed within the supercooled liquid to estimate properties related to nucleation (e.g., $n_c$, $\gamma$, $D$). In this manner, the method bypasses the challenge of nucleating the crystal from the liquid, but adds several additional assumptions about the nature of the nuclei. 

At moderate supercooling, the sampled critical nuclei are relatively small and surface-dominated. Employing such a small critical nucleus size in seeding would exacerbate the effect of errors in the definition of the nucleus size, $n$, on estimating the nucleation rate.\cite{Espinosa_Seeding_2016,Zimmermann_NaCl_2018,Jiang_Forward_2018,Sanz2021PCCP} To address this issue, Espinosa et al. demonstrated an alternative seeding method that is less sensitive to the nucleus size definition.\cite{Espinosa_Seeding_2016} Rather than growing seeds at the target (moderate) supercooling, we can simulate larger seeds at several lower supercoolings. We can estimate $\gamma$ at each lower supercooling using Eq.~\ref{eq:gamma-seed2}, obtained from rearranging Eq.~\ref{eq:critsize}.
\begin{equation}
\gamma = \bigg(\frac{3}{32\pi}\bigg)^{1/3} n_c^{1/3} \rho_\text{s}^{2/3} |\Delta \mu|.
\label{eq:gamma-seed2}
\end{equation}
Then $\gamma$ can be extrapolated from low supercooling to the target moderate supercooling. Once $\gamma$ is obtained at the condition of interest, $n_c$ is no longer directly evaluated with simulations but rather estimated from Eq. \ref{eq:critsize}. This prevents errors in $n$ from having such a large effect on the estimated nucleation rate. In this scheme, we first determined the critical temperature for seeds with $n>500$ at conditions of low supercooling ($0.85T_m-0.95T_m$). Once we obtained $\gamma$ values at these temperatures, a linear fit of $\gamma$ vs $T$ was constructed and extrapolated to the temperature of our RETIS simulations (Fig. S5). The value of $\gamma$ obtained at 0.78$T_m$ was then used along with $|\Delta \mu|$ and $\rho_\text{s}$ (see Section \ref{sec:CNT}) in Eq. \ref{eq:critsize} to determine $n_c$. With $n_c$, $|\Delta \mu |$, and $\rho_l$ determined, the only remaining unknown term in the rate equation (Eq. \ref{eq:seed-rate}) is $D$, which can be calculated from $dn/dt$ following the procedure from Ref. \citenum{Zimmermann_Nucleation_2015}. Full implementation details are provided in the SI.

\subsection{Reaction Coordinate Identification}

We employed the maximum likelihood estimation (MLE) approach, as described in Ref.~\citenum{Peters_Obtaining_2006}, to identify an optimized reaction coordinate from a set of physically meaningful order parameters (OPs). The committor probability, $p_B$, is widely regarded as the ideal reaction coordinate due to its ability to quantify a configuration’s likelihood of reaching the product state before the reactant. However, its statistical nature limits its direct interpretability for mechanistic insights.\cite{Rogal2020JCP,Peters:16:AnnRev} Instead, the $p_B$ is used to evaluate the quality of trial reaction coordinates by assessing the reaction coordinate correlation with the committor. Using the MLE framework,\cite{Peters_Obtaining_2006} we determined the combination of OPs that best approximated the committor and used it to characterize nucleation pathways. Implementation details are outlined in Section S6 of the SI.

\subsection{LeaPP: Learning Pathways to Polymorphs}

LeaPP is an analysis method that accounts for and characterizes nucleation pathways based on the time evolution of particle paths in an unsupervised manner.\cite{SarupriaJCTC2025} Below, we provide a brief overview of the LeaPP implementation. For a more detailed description and sample code, refer to Ref.~\citenum{SarupriaJCTC2025} and~\citenum{SarupriaLeaPPZenodo}, respectively. 

LeaPP includes three key steps. In step 1, we aim to learn a low-dimensional representation of each particle in the solid nuclei from all nucleation trajectories. We refer to this low-dimensional space as the latent space. The latent space was constructed by using a variational autoencoder (VAE)\cite{HigginsICLR2017} that was trained on each particle's local environment. This local environment was characterized by a set of Steinhardt bond order parameters (BOPs)\cite{SteinhardtPRB1983}, $q_l$, where $l=\{2,4,6,8\}$. Once the latent space was constructed, we projected each particle's time evolution through the VAE to the latent space to obtain ``particle paths''. Further details of the particle selection criteria, local descriptors, and the VAE architecture for the 7--6 and 12--6 system are described in Ref.~\citenum{SarupriaJCTC2025} and the SI, respectively. 

In step 2, the particle paths in the latent space were clustered using hierarchical agglomerative clustering based on their dynamic time warping (DTW)\cite{ChibaITASS1978,KeoghDMKD2017} pairwise distances. The underlying principle is that particles exhibiting similar evolution, indicative of comparable local environments, are likely to share the same identity and consequently, be grouped within the same cluster. This process yields distinct particle path clusters, which serve as labels for the particles assigned to each respective cluster. The resulting particle path clusters for both systems are denoted with Greek symbols (e.g. $\alpha$/$\beta$/$\gamma$/$\delta$). Correspondingly, particles which are assigned to the $\alpha$/$\beta$/$\gamma$/$\delta$ cluster would be referred to as $\alpha$/$\beta$/$\gamma$/$\delta$ particles, respectively. These particle assignments allow us to identify the roles of participating particles in nucleation, and are used in the next step for clustering nucleation trajectories. 

In step 3, we characterized the nucleation trajectories into groups of nucleation pathways. To perform this characterization, first, a normalized distribution of particle types was constructed for each nucleation trajectory. The distribution counts the total number of different particle types from step 2 in the largest solid nucleus\cite{ten_Wolde_Numerical_1996, tenWolde:05:PRL} throughout the frames of a trajectory.\cite{SarupriaJCTC2025} Once each nucleation trajectory had a representative distribution, the pairwise differences between all nucleation trajectory distributions were computed to obtain a difference metric for hierarchical agglomerative clustering of the nucleation trajectories. The resulting trajectory clusters from this last step constitute nucleation pathways of the system. 

\section{Results and Discussion}
To elucidate the differences between nucleation in the 7–6 and 12–6 systems, we compare nucleation rates obtained from RETIS. We apply CNT with seeding to extract and interpret the thermodynamic contributions to the nucleation barrier and corresponding rate. Additionally, we characterize the nucleation mechanism of each system by determining the reaction coordinate and nucleation pathways.

\subsection{Nucleation Rates of the 7--6 and 12--6 Systems}\label{sec:nuc_rate}

\begin{figure}[h!]
\begin{center}
\includegraphics[width=\linewidth]{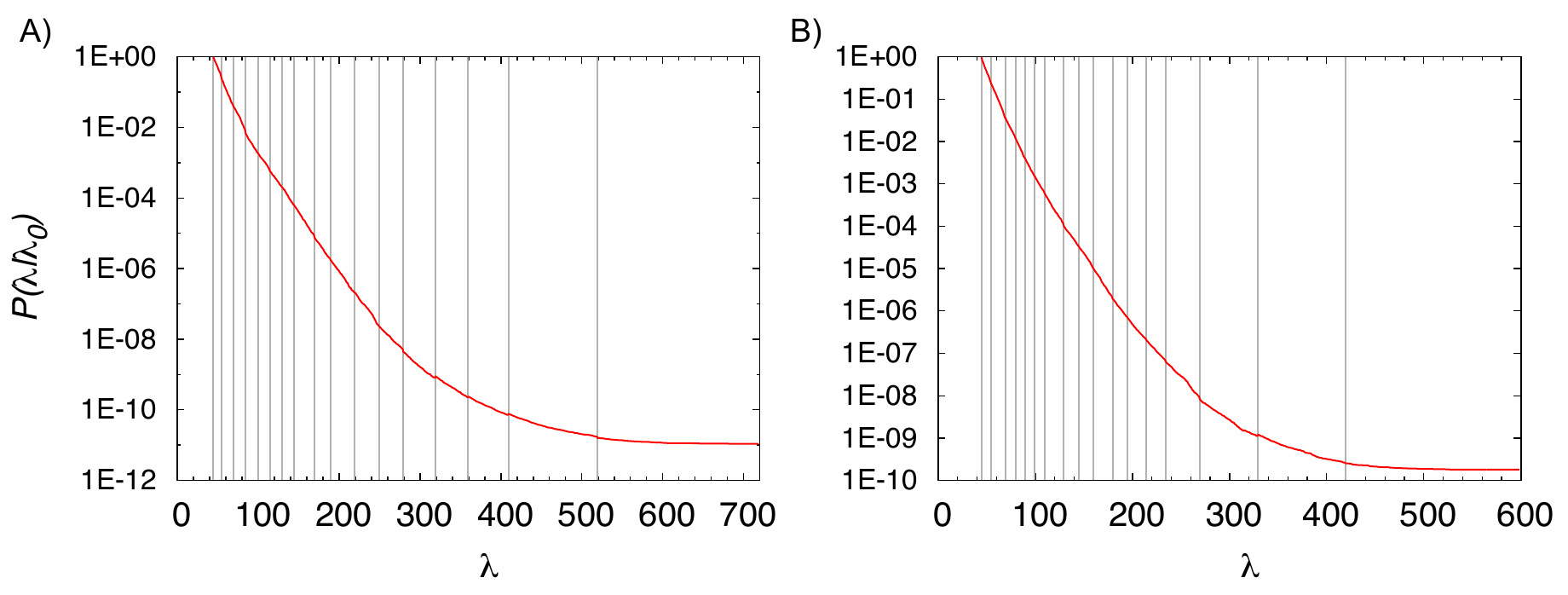}
\caption{Crossing probability histogram, $P(\lambda|\lambda_0)$, constructed from WHAM. The vertical lines represent each interface ensemble, $\lambda_i$. The red line follows the probability of reaching some interface value ($\lambda$) given that it starts from the first interface. The physical interpretation of this plot is: (A) for the 7--6 system, the probability of reaching a crystal size of 720 particles, starting from a crystal size of 45, is $~1 \times 10^{-11}$. (B) For the 12--6 system, the probability of reaching a crystal size of 600 particles, starting from a crystal size of 45, is $~2 \times 10^{-10}$.}
\label{fig:wham}
\end{center}
\end{figure}

In RETIS, the nucleation rate is given by:

\begin{equation}
    J = \frac{k_{AB}}{V_{avg,liq}} = \frac{\Phi_0\times P(\lambda_N | \lambda_0)}{V_{avg,liq}}
\label{eq:nuc_rate}
\end{equation}
where $\Phi_0$ is the flux of trajectories crossing the first interface ($\lambda_0$) from the liquid basin, $P(\lambda_N|\lambda_0)$ is the probability of a trajectory crossing the final interface from the first interface, and $V_\text{avg, liq}$ is the average volume of the simulation box containing the liquid phase.\cite{vanErp:07:PRL, vanErp:17:JCP} The flux $\Phi_0$ is calculated from the average path lengths of the first interface ensemble\cite{vanErp:17:JCP} and the volume term is computed directly from the simulation. $P(\lambda_{N}|\lambda_0)$ was obtained from the final value of the crossing probability histogram shown in Fig.~\ref{fig:wham}. This overall crossing probability histogram was obtained from stitching the local crossing probability histogram of each interface (Fig. S4) via the weighted histogram analysis method (WHAM).\cite{Swendsen1989PRL, SarupriaJCP2022} The uncertainty in $P(\lambda_N|\lambda_0)$ value was calculated from 500 bootstrap samples, each consisting of 440 paths randomly sampled with replacement from each interface ensemble.\cite{SarupriaJCP2022}

Nucleation rates were calculated using Eq. \ref{eq:nuc_rate}. Despite the different interaction potentials, the nucleation rates for both systems are comparable (Table \ref{tab:rate-calc}). Although the 7--6 system has a slightly lower crossing probability, it has a larger flux and a smaller system volume. Together, these factors produce a nucleation rate for the 7--6 system that is only one order of magnitude slower than the 12--6 system. In the context of nucleation rates, a difference of about one order of magnitude is negligible.\cite{Sosso_Crystal_2016, Barbora2015EPJ}
\begin{table}[h!]
\linespread{0.9}\selectfont\centering
\caption{Rate calculations for the 7--6 and 12--6 systems. The volume is reported in $\sigma^3$.}
\label{tab:rate-calc}
\begin{center}
\setlength{\tabcolsep}{10pt}
\begin{tabular}{@{} l  c c @{}} 
\toprule
 & 7--6 & 12--6 \\ 
\midrule
$\log_{10}{\Phi_0}$ & $-0.95$ & $-1.29$ \\
$\log_{10}{P(\lambda_N|\lambda_0)}$ & $-10.9(0.4)$ & $-9.7(0.4)$ \\
$V_\text{avg,liq}$ & $6888.1$ & $8368.1$ \\
$\log_{10}{J}$ & $-15.8(0.4)$ & $-14.9(0.4)$\\
\bottomrule
\end{tabular} 
\end{center}  
\end{table}

Given the fundamental differences in the interaction potentials of the two systems, the similarity in their nucleation rates is surprising and not immediately intuitive. This observation implies that their free energy barriers are comparable, possibly due to offsetting changes in the critical cluster size or interfacial tension. This raises interesting questions about the sensitivity of nucleation kinetics to the potential form. We leverage the CNT framework to examine this further and unpack the underlying thermodynamic and kinetic contributions. Through seeding, we estimated all relevant properties for predicting the nucleation barrier and rate according to CNT (Table \ref{tab:thermoProperties}). 
Under the conditions used in RETIS, the kinetic prefactor, given by the product $\rho_{\text{l}} D Z$, is of the order of unity---meaning that the thermodynamic nucleation barrier is the dominant factor controlling the nucleation rate. Assuming spherical nuclei, Eqs. \ref{eq:critbarrier} and \ref{eq:critsize} indicate that the barrier height depends on $\Delta\mu$, $\rho_{\text{s}}$, and $\gamma$. From Table \ref{tab:thermoProperties}, the product $\rho_{\text{s}}|\Delta\mu|$, which represents the driving force per unit volume, is nearly identical for both potentials. However, since $\gamma_{7-6}$ is slightly larger than $\gamma_{12-6}$, the critical nucleus size for the 7--6 system is correspondingly larger. This leads to a higher energy barrier and a lower nucleation rate for the 7--6 system, as is observed from the RETIS rate.
\begin{table}[h]
\linespread{0.9}\selectfont\centering
\caption{Properties of 7--6 and 12--6 systems at $p=5$, $T=0.78T_m$ calculated from seeding simulations. $\gamma$ was estimated by extrapolating from higher temperature seeding simulations. Uncertainties at 95\% confidence level are given in parentheses.}
\label{tab:thermoProperties}
\begin{center}
\setlength{\tabcolsep}{3pt}

\begin{tabular}{@{} l  c  c  c  c  c  c  c  c c c @{}} \toprule
Potential & $\rho_{\text{s}}$ & $\rho_{\text{l}}$ & $|\Delta\mu|$ & $\rho_{\text{s}}|\Delta\mu|$ & $\gamma$ & $n_c$ & $D$ & $\Delta G_{\text{c}}$ & $\log_{10}{J_\text{seed}}$&$\log_{10}{J_\text{RETIS}}$\\
\midrule
7--6  &  1.254  &  1.189  &  0.242  &  0.304  &  0.57(0.02)  &  287  & 221.0 & 34(4) & -17.5(1.9)&-15.8(0.4)\\
12--6  &  1.054  &  0.979  &  0.289  &  0.305  &  0.53(0.01)  &  182  & 108.4 & 26(2) & -13.6(1.0) & -14.9(0.4)\\
\bottomrule

\end{tabular} 
\end{center}  
\end{table}

The difference in nucleation rates between the two systems, as predicted by RETIS and seeding, is small when considering the associated uncertainties. On the other hand, the rate discrepancy between the two methods is a more interesting observation. Although RETIS and seeding predicted the same trend for both the 7--6 and 12--6 systems, there is approximately two orders of magnitude difference between the predictions from both methods. We suspect that, for the 12--6 system, this discrepancy stems from the use of a perfect FCC seed in the seeding approach. This likely leads to an underestimated critical nucleus size and, consequently, an overestimated nucleation rate. Similar overestimations have been observed in NaCl nucleation, where seeding produced higher rates compared to FFS.\cite{Zimmermann_NaCl_2018, Jiang_Forward_2018} However, modifying the OP definition in seeding has been shown to reconcile this discrepancy, yielding nucleation rates comparable to those obtained from FFS.\cite{Sanz2021PCCP, Dijkstra2024JCP} For the 7--6 system, the source of the discrepancy is less apparent. We hypothesize that this discrepancy also relates to deviations of the critical nuclei structure from the pure FCC structure, as assumed in our seeding approach.

\subsection{Nucleation Mechanism of the 7--6 and 12--6 System}\label{sec:nuc_mech}
Although the nucleation rates between the two systems are comparable, this does not necessarily imply that the underlying nucleation mechanisms are equivalent. While the rate quantifies the height of the nucleation barrier, it offers no direct insight into the structural pathway by which the system traverses the barrier. Given the distinct interaction potential of the two systems, a key question arises: Do similar rates imply similar pathways? To explore this, we estimated their respective optimal reaction coordinate and determined the factors that drive the nucleation event. We also applied LeaPP to the ensembles of reactive trajectories to characterize and compare the nucleation pathways.

\subsubsection{Reaction Coordinate}

Using MLE, we found that the combination of the largest nucleus size (n$_\text{tf}$) and the average crystallinity of the nucleus (Q$_\text{6,tf}^\text{cl}$) based on $q_{6}$\cite{ten_Wolde_Numerical_1996, tenWolde:05:PRL} serve as the optimal reaction coordinate for both the 7--6 and 12--6 system (Table S3 and S4). We confirmed the validity of this reaction coordinate via the histogram test, following the protocol from Ref.~\citenum{Peters_Obtaining_2006} (see SI Section S6.3). We estimated the committor by projecting the reweighted path ensemble (RPE) from RETIS onto these two OPs in discretized bins using the scheme from Rogal et al.~\cite{Bolhuis:10:JCP}. Each configuration is assigned a committor value corresponding to the bin to which it belongs.

\begin{figure}[h!]
    \centering
    \includegraphics[width=\linewidth]{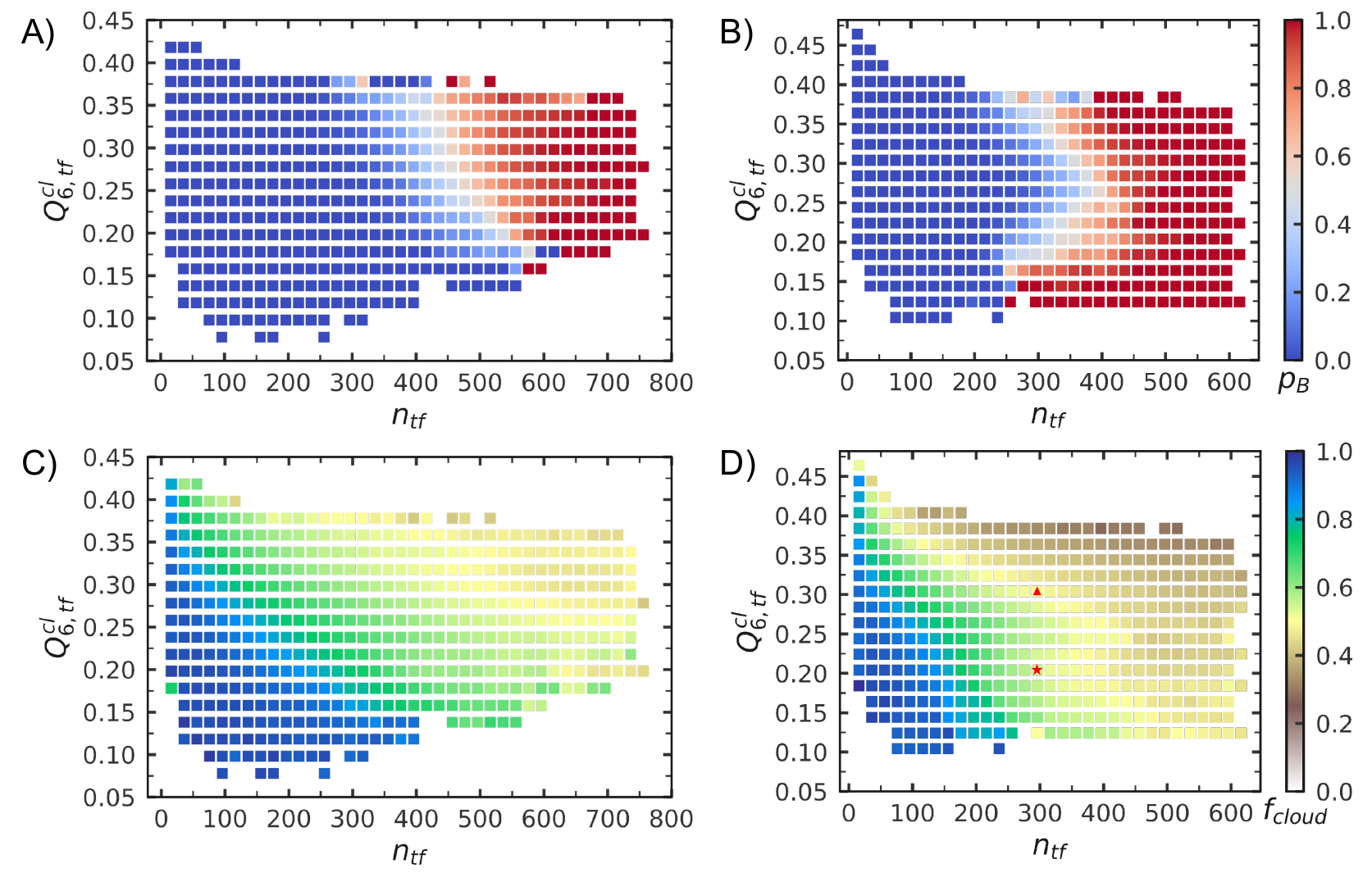}
    \caption{$p_B$ projection of configurations from (A) 7--6 and (B) 12--6 system onto n$_\text{tf}$ (x-axis) and Q$_\text{6,tf}^\text{cl}$ (y-axis). The plot is colored by the committor value from 0 to 1. Projection of the average fraction of cloud particles (f$_\text{cloud}$) along n$_\text{tf}$ and Q$_\text{6,tf}^\text{cl}$ for the (C) 7--6 and (D) 12--6 system. The red triangle and star on panel D signify the configurations where n$_\text{tf} = 300$ and Q$_\text{6,tf}^\text{cl}\approx 0.20$ (star)/Q$_\text{6,tf}^\text{cl}\approx 0.30$ (triangle). The color map reflects the values of f$_\text{cloud}$, which increases from 0 to 1. Details of this OP calculation are described in SI Section S5.}
    \label{fig:pbproj}
\end{figure}

The committor projection in Fig. \ref{fig:pbproj} indicates that both systems progress in the direction of increasing n$_\text{tf}$ size and crystallinity. Although the optimized reaction coordinate is identical for both systems, we observed distinct differences in their transition region configurations. In the 7--6 system, the transition state configurations range from smaller and more crystalline configurations to those that are larger and less crystalline (Fig. \ref{fig:pbproj}A). In contrast, the 12--6 system exhibits a transition region that is characterized by an elbow shape (Fig. \ref{fig:pbproj}B). For instance, transition configurations with Q$_\text{6,tf}^\text{cl}>0.25$ (i.e., the upper elbow region) range from small and more crystalline nuclei to those that are larger and less crystalline. However, for configurations with Q$_\text{6,tf}^\text{cl}<0.25$ (i.e., the lower elbow region), the trend reverses. These configurations range from those that are larger and more crystalline to those that are smaller and less crystalline. Compared to previous studies of 12--6 nucleation at slightly higher supercooling (0.75$T_m$) from Moroni \textit{et al.}\cite{Moroni_Interplay_2005} and Beckham \textit{et al.}\cite{Beckham_Optimizing_2011}, our findings show an agreement and extension. The upper elbow transition region aligns with the transition region reported in Ref.~\citenum{Moroni_Interplay_2005} and \citenum{Beckham_Optimizing_2011}. The lower region of the elbow-like transition region reveals a portion of the 12--6 transition region that was not sampled in earlier studies. Nuclei in this region contain a greater fraction of HCP particles compared to FCC (see Fig. S10). To further validate whether the configurations from the lower elbow region are indeed in the transition state,  we performed the committor histogram test\cite{Peters_Obtaining_2006} on configurations sampled exclusively from this region (see Fig. S9). The histogram peaks at $\sim$0.5, confirming that the configurations from the lower elbow region are in the transition state. We hypothesize that this region was not sampled in previous studies for the following reasons. Beckham \textit{et. al.} performed aimless shooting with an initial configuration obtained from an FCC seed placed in the liquid. It is possible that this resulted in biasing the configurations to sample nuclei with higher FCC content. Moroni \textit{et. al.} performed simulations in the NPH ensemble, which allows for local annealing of the growing crystal. This results in higher crystallinity of the nuclei, as observed in Refs. \citenum{Kusalik2013JPCB} and \citenum{Guo2016PCCP}, where the influence of temperature and pressure coupling on hydrate crystallinity was investigated.
This indicates that differences in sampling approaches---particularly those that may preferentially explore more crystalline FCC nuclei---could contribute to the lower elbow region being less frequently sampled.

This elbow-like trend in the 12--6 transition region is noteworthy, as this observation suggests that two critical nuclei with the same size and different crystallinity have an equal probability of overcoming the barrier. For example, at n$_\text{tf}$ $\approx 300$, both low (Q$_\text{6,tf}^\text{cl}\approx0.20$) and high crystallinity (Q$_\text{6,tf}^\text{cl}\approx0.30$) critical nuclei have about the same committor probability of 0.5. We hypothesize that the differing role of crystallinity in the transition region of both systems may be linked to the variations of cloud particles within the critical nuclei. According to Ref.~\citenum{Bolhuis2011PRL}, cloud particles are surface particles of the n$_\text{tf}$ nucleus that possess a structural order that is intermediate between liquid and bulk solid. Since the crystallinity OP measures how different the local environment of solid particles in the nucleus is from bulk liquid, it is likely inversely correlated with the fraction of cloud particles (f$_\text{cloud}$) that exist within the n$_\text{tf}$ nucleus.

To explore this, we calculated the average f$_\text{cloud}$ in each bin of the committor projection for the 7--6 and 12--6 systems (Fig. \ref{fig:pbproj}C-D). For the 12--6 system at n$_\text{tf} = 300$, we observe that critical nuclei that contain a smaller fraction of f$_\text{cloud}$ (triangle in Fig.~\ref{fig:pbproj}D) exhibit a higher value of crystallinity, while those with a higher fraction of f$_\text{cloud}$ have a lower crystallinity value (star in Fig.~\ref{fig:pbproj}D). This result supports the inverse correlation between crystallinity and f$_\text{cloud}$, and hints at the possibility that the cloud particle content of the n$_\text{tf}$ nucleus may underlie the observed differences in crystallinity for nuclei of equal size and committor probability.

\begin{figure}[h!]
    \centering
    \includegraphics[width=0.7\linewidth]{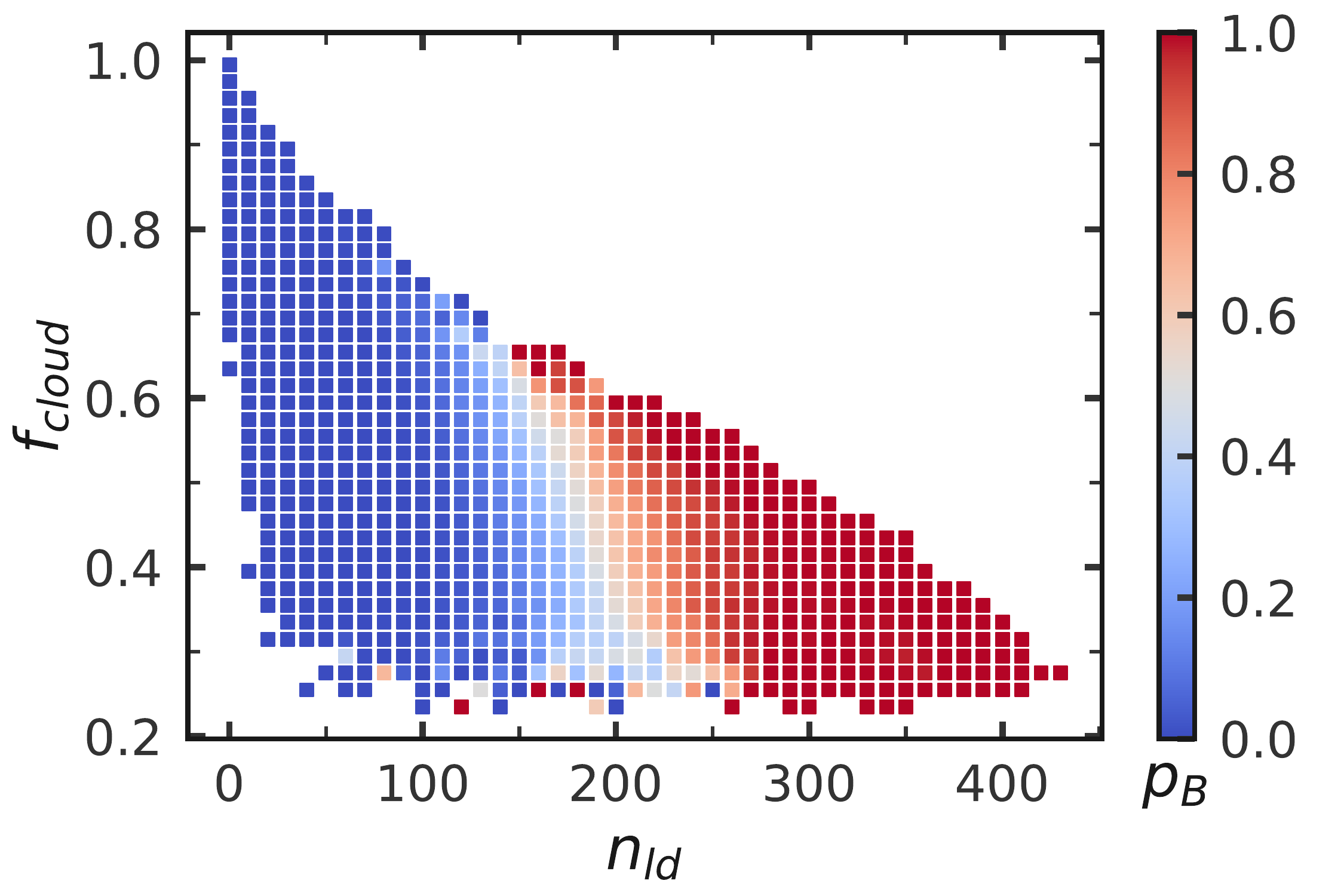}
    \caption{$p_B$ projection of configurations from the 12--6 system onto n$_\text{ld}$ (x-axis) and f$_\text{cloud}$ (y-axis). The plot is colored by the committor value from 0 to 1.}
    \label{fig:12-6_newRC}
\end{figure}

Motivated by this, we reevaluated the reaction coordinate using the MLE algorithm, now including f$_\text{cloud}$ as an OP. We found that the combination of f$_\text{cloud}$ with a stricter largest nucleus size OP---n$_\text{ld}$, as defined by Lechner and Dellago~\cite{Lechner_Accurate_2008}---yields a more effective reaction coordinate for the 12--6 system compared to n$_\text{tf}$Q$_\text{6,tf}^\text{cl}$ (Fig.~\ref{fig:12-6_newRC}). Unlike the n$_\text{tf}$, which relies on the standard $q_6$ BOP to assess particle solidity, n$_\text{ld}$ employs the averaged local BOP, $\overline{q}_6$, resulting in a more stringent classification of solid-like particles.\cite{Lechner_Accurate_2008} Consequently, n$_\text{ld}$ tends to be smaller than n$_\text{tf}$, and can be viewed as representing the core of the n$_\text{tf}$ nucleus, as suggested in Ref.~\citenum{Bolhuis2011PRL}. As such, f$_\text{cloud}$ is also defined as the surface particles between n$_\text{ld}$ and n$_\text{tf}$ given by $\frac{n_\text{tf} - n_\text{ld}}{n_\text{tf}}$.

The combination of n$_\text{ld}$ and f$_\text{cloud}$ as a representation of the reaction coordinate provides a more nuanced understanding of the nucleation mechanism and helps resolve the earlier observation of variability in crystallinity among critical nuclei of the same n$_\text{tf}$ size. This discrepancy in crystallinity can now be interpreted in terms of the internal structure of the n$_\text{tf}$ nucleus---specifically, the relative size of its crystalline core (n$_\text{ld}$) and the extent of its surface cloud particle region (f$_\text{cloud}$). At a fixed n$_\text{tf}$ size, nuclei with larger n$_\text{ld}$ values contain fewer cloud particles, leading to higher crystallinity. Conversely, nuclei with smaller n$_\text{ld}$ have a higher fraction of cloud particles, and thus lower crystallinity. Despite these differences in the internal structure of n$_\text{tf}$, both types of critical nuclei---at the same n$_\text{tf}$ size---can occupy the same region of the committor landscape, as seen in Fig.~\ref{fig:12-6_newRC}.

Overall, the reaction coordinates obtained for both the 12--6 system and 7--6 system incorporate two key descriptors: the size of the nucleus and a measure of its crystallinity. This highlights the interplay between crystallinity and nucleus size as a key feature of the nucleation process.

\subsubsection{Critical Nuclei Composition}

\begin{figure}[h!]
    \centering
    \includegraphics[width=\linewidth]{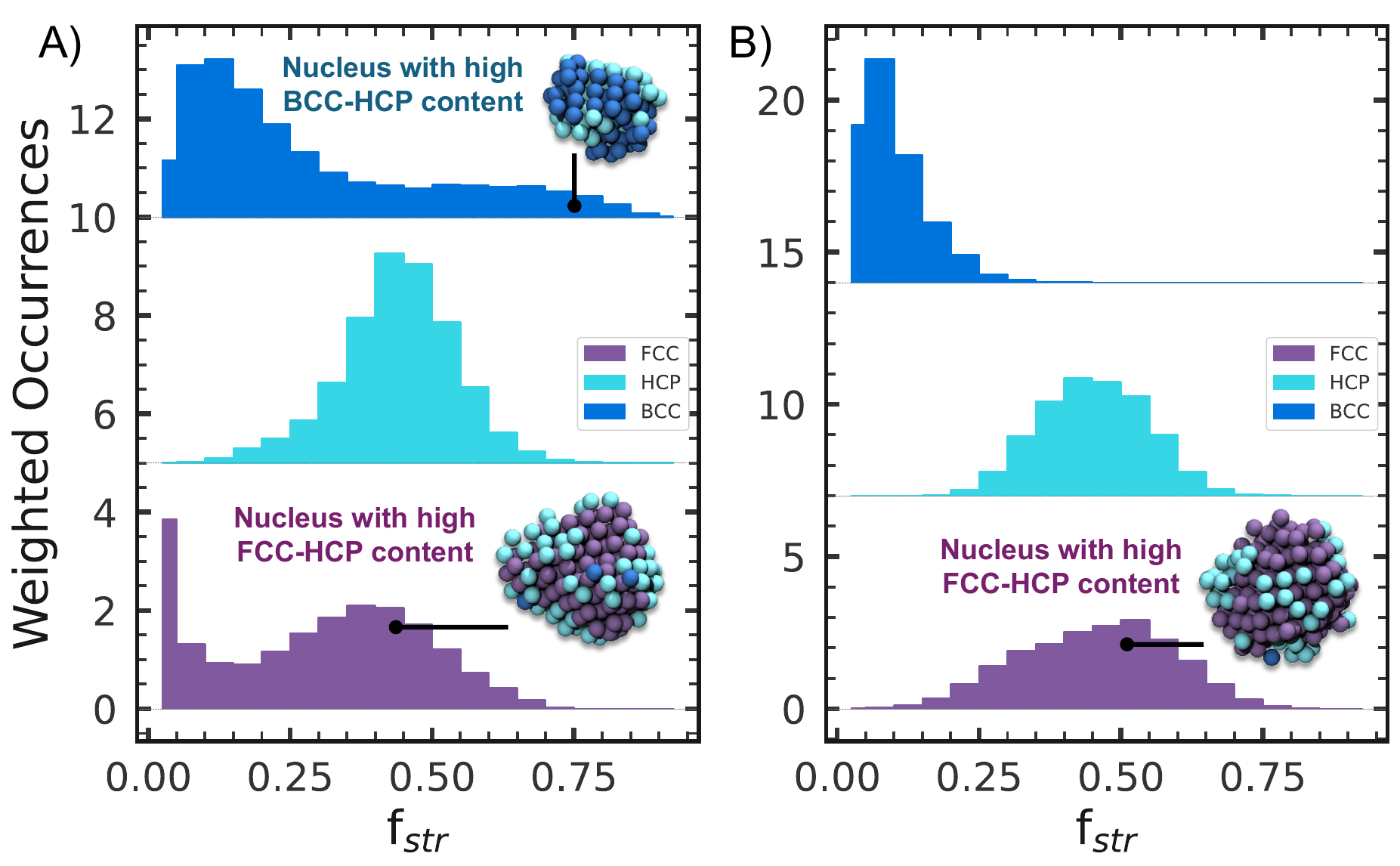}
    \caption{The distribution of critical nucleus composition for (A) 7-6 and (B) 12-6 system. We define critical nucleus configurations as those with $0.45 \leq p_B \leq 0.55$. The x-axis describes the fraction of each composition as shown by the colors: blue -- BCC, cyan -- HCP, purple -- FCC. The y-axis accounts for the weighted contributions of each configuration (and the curves are shifted for visual clarity). Each configuration weight is calculated from the reweighted path ensemble weight of each trajectory.}
    \label{fig:ts-comp}
\end{figure}

The nuanced difference in the reaction coordinate between the 7--6 and 12--6 systems inspired us to further investigate the composition of their transition state configurations (i.e., critical nuclei). We defined the composition of each configuration in terms of the fraction of BCC (f$_\text{BCC}$), HCP (f$_\text{HCP}$) and FCC (f$_\text{FCC}$) within the largest solid nucleus. Figure~\ref{fig:ts-comp} illustrates the distribution of f$_\text{BCC}$, f$_\text{HCP}$, and f$_\text{FCC}$ from all critical nuclei configurations. For the 7--6 system in Fig. \ref{fig:ts-comp}A, there exist two composition profiles: (i) a nucleus that is more dominantly BCC throughout, and (ii) a nucleus that is comprised of an FCC core with an HCP interface. As for the 12--6 system, only nuclei with an FCC core and an HCP interface exist (Fig. \ref{fig:ts-comp}B). This result is consistent with previous studies that have shown that softer potential increases the stability of the BCC structure.\cite{Likos:01:PhysRep,vanblaaderen2003Nature,Lowen2004PRL,Giaquinta2005JCP,Yethiraj_Tunable_2007,Delhommelle:07:JCP,Delhommelle:07:JPCB,Boles_Self_2016,Li_Assembly_2016,Schwerdtfeger2021PRE} More importantly, this demonstrates that 12-6 and 7-6 systems may have different structural nucleation pathways. 

\subsubsection{Particle Path Clusters}

The difference in the composition of critical nuclei structures of both systems alludes to the possibility of a difference in their nucleation pathways. We utilized LeaPP (Learning Pathways to Polymorphs)~\cite{SarupriaJCTC2025} to further investigate the nucleation pathways of each system. LeaPP, as described in Ref.~\citenum{SarupriaJCTC2025}, uses the time evolution of particles participating in the nucleation event to elucidate nucleation pathways in an unsupervised manner. 

The latent space projection from the VAE (step 1 of LeaPP) indicates that the liquid particle points are distinguished from the solid particles for both systems (Fig. S12-S13). Furthermore, the FCC particles are distinguished from BCC and HCP particles. Overall, the result demonstrates that the VAE learns the local environment of each particle and constructs a reasonable latent space representation. 

\begin{figure}[h!]
    \centering
    \includegraphics[width=\linewidth]{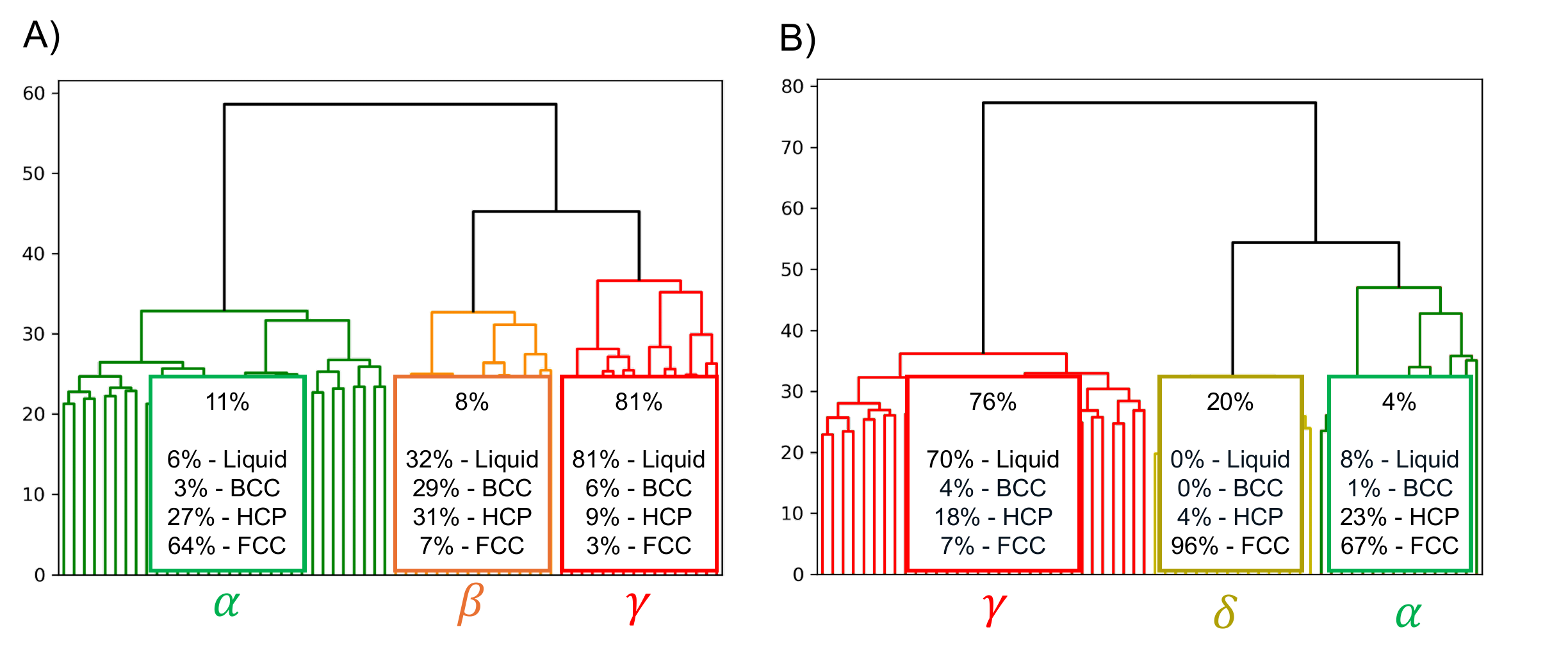}
    \caption{Hierarchical agglomerative clustering of (A) 7--6 and (B) 12--6 particle paths. Each colored region of each dendrogram represents a cluster of particle paths. The 7--6 particle path clusters consist of $\alpha$, $\beta$ and $\gamma$. The 12--6 particle path clusters consist of $\alpha$, $\gamma$ and $\delta$. The percentage at the top center reflects the fraction of particle paths that belong to each cluster. The amount of liquid, BCC, FCC, and HCP in each particle path cluster is calculated based on the structural identities ($\overline{q_4}-\overline{q_6}$ values) of the particles in the final frame. These values represent the phase composition of particles within each cluster, that is, the percentage of the four categories will sum to 100 for each particle path cluster.}
    \label{fig:ppc}
\end{figure}

The time evolution of all nucleating particles was projected through the VAE to obtain particle paths in the latent space. Pairwise differences between particle paths were computed via DTW and used as inputs for particle path clustering (step 2 of LeaPP). Using the selected cutoff criteria, Fig. \ref{fig:ppc} shows that three distinct particle path clusters emerge for each system. We interpret each particle path cluster according to the identity of the particles that are clustered. The particle identity was defined according to their $\overline{q_4}$ and $\overline{q_6}$ values at the last frame of the nucleation trajectory. In the 7--6 system, the clusters are labeled $\alpha$, $\beta$, and $\gamma$ (Fig. \ref{fig:ppc}A), and in the 12--6 system, the clusters are labeled $\alpha$, $\delta$, and $\gamma$ (Fig. \ref{fig:ppc}B). The cluster labels $\alpha$ and $\gamma$ were utilized across both systems to reflect the two systems' similar particle path cluster compositions, thereby drawing attention to differing clusters---$\beta$ in the 7--6 system and $\delta$ in the 12--6 system.

In the 7--6 system, the $\alpha$ particle path cluster mainly contains FCC particles, and a small amount of HCP particles. The $\beta$ particle path cluster features a balanced composition of liquid, HCP, and BCC particles, while the $\gamma$ particle path cluster is dominated by liquid particles. The 12--6 particle path clusters (Fig. \ref{fig:ppc}B) also show a similar pattern for the $\alpha$ and $\gamma$ clusters. Notably, the 12--6 does not have any particle path cluster that contains more than 5\% BCC in its cluster definition. In place of the $\beta$ cluster seen in the 7--6 system, the $\delta$ cluster in the 12--6 system primarily contains FCC particles. This distinction aligns with the observation from Fig. \ref{fig:ts-comp}, where the softer potential has BCC-dominant and FCC-dominant critical nuclei, while 12-6 system has only FCC-dominant nuclei. These observations further support the hypothesis that the two systems may have different nucleation pathways. 

\subsubsection{Nucleation Pathways}
Using the particle path clusters, we generated a bar graph for each nucleation trajectory to capture the distribution of particles (i.e., $\alpha$, $\beta$, $\gamma$, $\delta$ particles) across all frames of the trajectory (Step 3 of LeaPP). We calculated the pairwise differences between these bar graphs to construct a distance matrix that quantifies the dissimilarity between each pair of trajectories. This matrix served as input for hierarchical agglomerative clustering, allowing us to group similar trajectories into nucleation trajectory clusters. These nucleation trajectory clusters were subsequently analyzed to identify and characterize distinct nucleation pathways. 

For comparison purposes, we chose a clustering cutoff such that both systems would have two nucleation trajectory clusters (see SI Fig. S14). This results in TC-S1 and TC-S2 nucleation trajectory clusters for the 7--6 system and TC-L1 and TC-L2 trajectory clusters for the 12--6 system. We performed radial composition analysis (RCA)~\cite{Moroni_Interplay_2005,SarupriaJCTC2025} on configurations extracted from each trajectory cluster of each respective system. One configuration from each trajectory within the trajectory cluster was randomly selected for the RCA if their $p_B$ value is 0.9 or higher. Details of the RCA calculation are described in Ref.~\citenum{SarupriaJCTC2025}. 

\begin{figure}[h!]
    \centering
    \includegraphics[width=\linewidth]{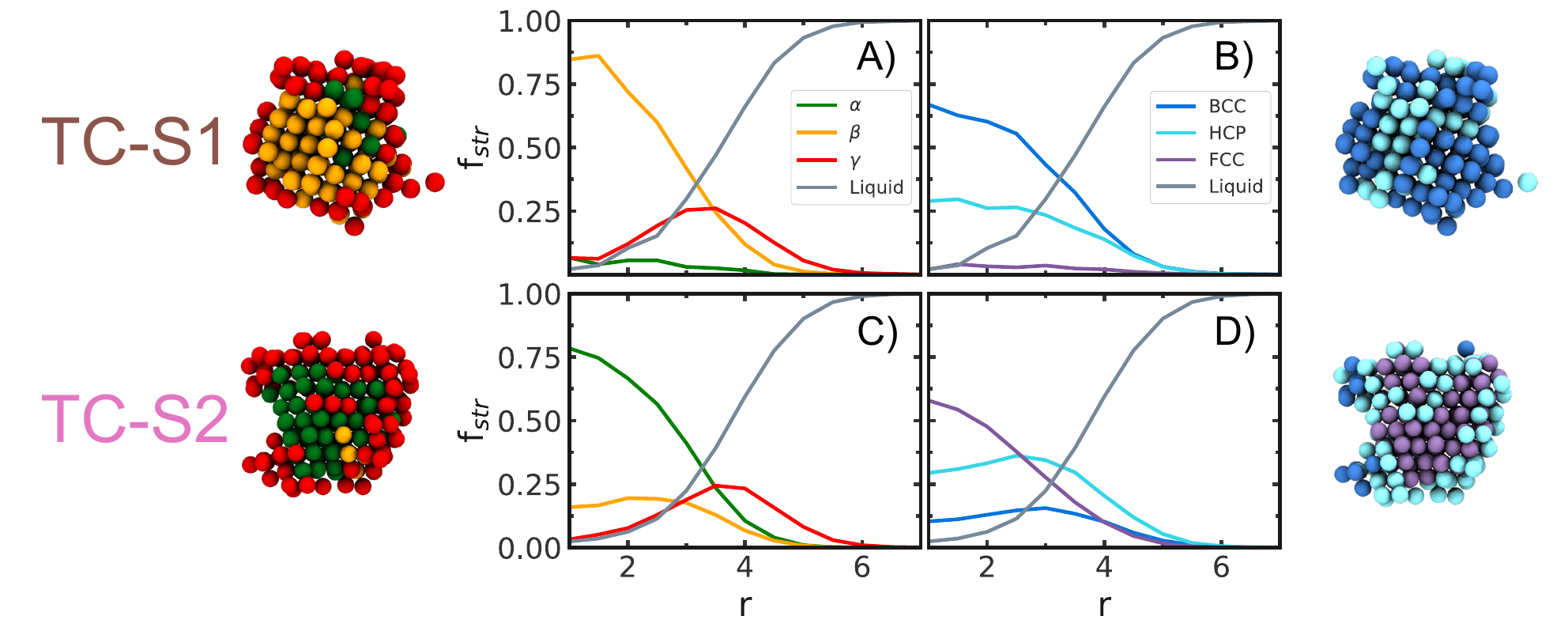}
    \caption{Radial composition profile of configurations from 7--6 nucleation trajectory clusters. (A-B) Radial composition profile of TC-S1 nucleation trajectory cluster. (C-D) Radial composition profile of TC-S2 nucleation trajectory cluster. Panels A and C reflect the radial profile where LeaPP labels are used. Panels B and D reflect the radial profile where particles are labeled based on their $\overline{q_4}$ and $\overline{q_6}$ values. r is the distance from the center of the nucleus core (units of $\sigma$). f$_{str}$ is the fraction of each composition as a function of the distance r.  Reprinted (adapted) with permission from Hall, S. W.; Minh, P.; Sarupria, S.\textit{ LeaPP: Learning Pathways to Polymorphs through Machine Learning Analysis of Atomic Trajectories. J. Chem. Theory Comput.} \textbf{2025}, 21 (8), 4121–4133. https://doi.org/10.1021/acs.jctc.5c00097. Copyright 2025 American Chemical Society.}
    \label{fig:rca76}
\end{figure}

\begin{figure}[h!]
    \centering
    \includegraphics[width=\linewidth]{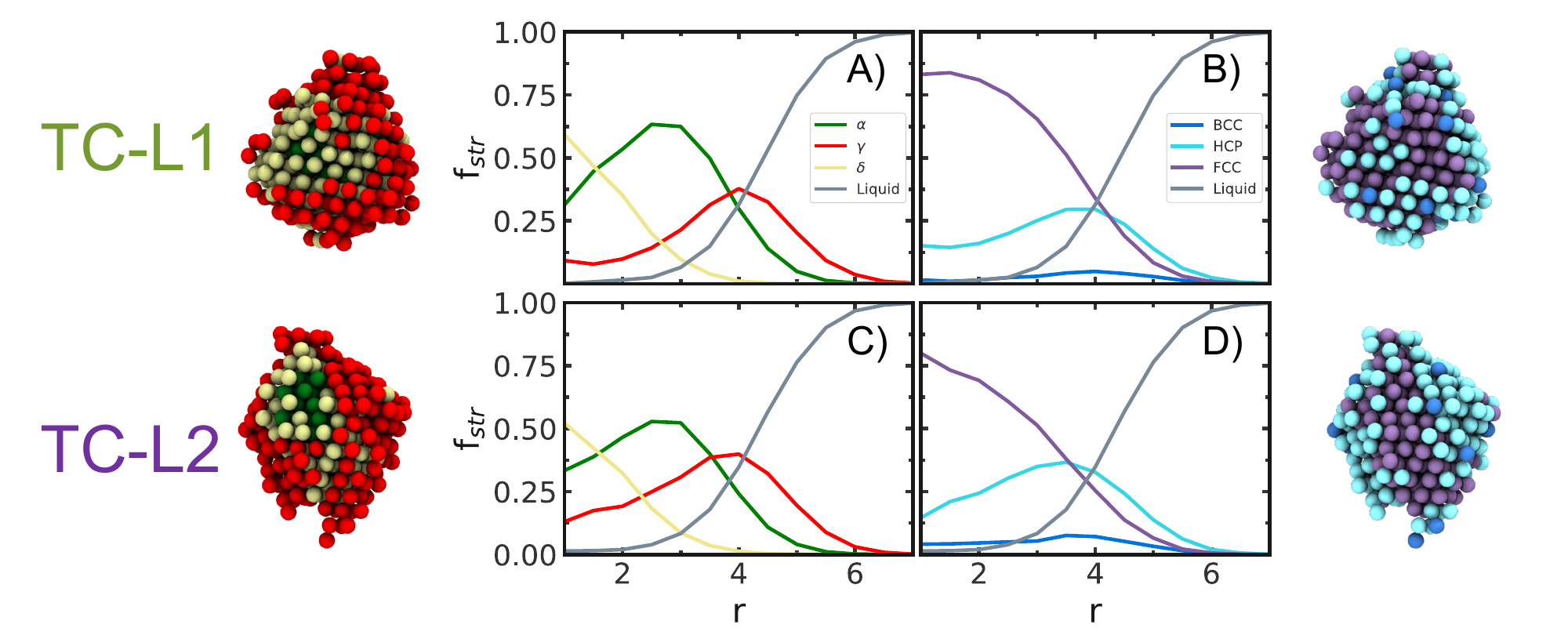}
    \caption{Radial composition profile of configurations from 12--6 nucleation trajectory clusters. (A-B) Radial composition profile of TC-L1 nucleation trajectory cluster. (C-D) Radial composition profile of TC-L2 nucleation trajectory cluster. Panels A and C reflect the radial profile where LeaPP labels are used. Panels B and D reflect the radial profile where particles are labeled based on their $\overline{q_4}$ and $\overline{q_6}$ values.}
    \label{fig:rca126}
\end{figure}

The RCA profiles for the trajectory clusters are shown in Fig.~\ref{fig:rca76} and ~\ref{fig:rca126} for the 7--6 and 12--6 systems, respectively. We performed RCA on the configurations using labels from both LeaPP's particle path clusters and $\overline{q_4}$ and $\overline{q_6}$ values. 
For the 7--6 system, TC-S1 and TC-S2 have different radial profiles (Fig.~\ref{fig:rca76}). In the LeaPP definition, TC-S1 nuclei have a dominant $\beta$ core followed by a $\gamma$ interface. In the $\overline{q_4}$ and $\overline{q_6}$ definition, these configurations have a mixture of BCC and HCP throughout the nucleus. As for TC-S2, these nuclei have an $\alpha$ core followed by a $\gamma$ interface, which in the $\overline{q_4}$ and $\overline{q_6}$ definition translates to nuclei with a dominant FCC core and a mixture of FCC-HCP interface.  
This result demonstrates that TC-S1 and TC-S2 nucleation trajectories could progress through two different pathways. 

In contrast to the 7--6, the 12--6 trajectory clusters, TC-L1 and TC-L2, do not have a difference in their radial profiles. TC-L1 and TC-L2 both have a mixture of $\delta$ and $\alpha$ particles in the core, followed by $\gamma$ particles in the interface. In the $\overline{q_4}$ and $\overline{q_6}$ definition, these nuclei are overall dominant in FCC content at the core and a mixture of FCC-HCP at the interface. The results indicate that, while the 7--6 system could nucleate through two different pathways, the 12--6 system exhibits only one nucleation pathway. 

\begin{figure}[h!]
    \centering
    \includegraphics[width=0.8\linewidth]{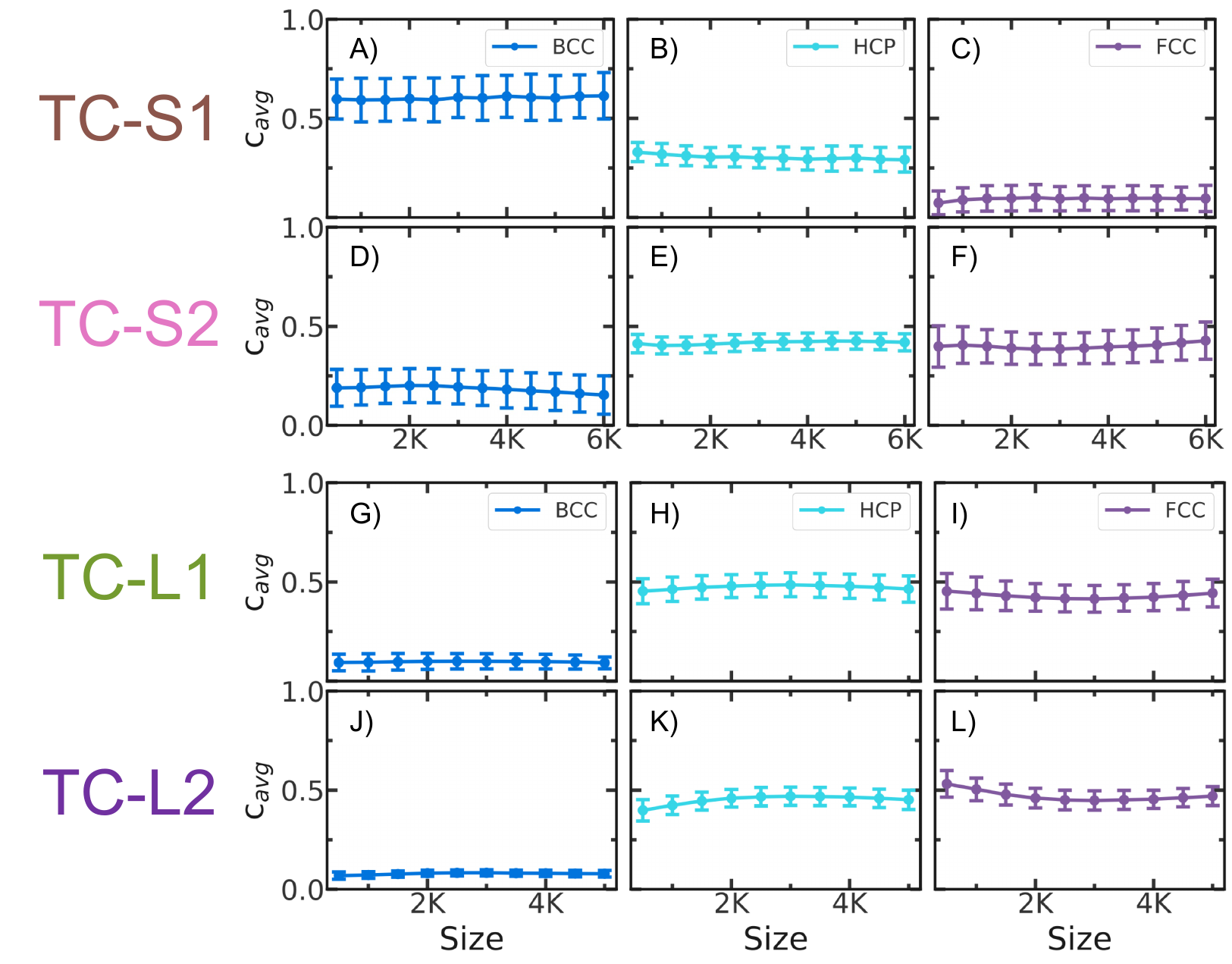}
    \caption{Average composition of (A-F) 7--6 and (G-L) 12--6 nucleation trajectories from each trajectory cluster in Fig. S14. The x-axis represents the average nucleus size of the configurations in the trajectories of each respective cluster. c$_\text{avg}$ represents the average composition of the nucleus at a specific size according to the three phases: BCC, HCP, and FCC. Blue lines represent the BCC composition, cyan lines represent the HCP composition and purple lines represent the FCC composition. The composition is calculated for trajectories in the TC-S1 cluster (A-C), the TC-S2 cluster (D-F), the TC-L1 cluster (G-I), and the TC-L2 cluster (J-L). Details of this calculation are described in Ref.~\citenum{SarupriaJCTC2025}. The error bars reflect the standard deviation of the average composition across all trajectories launched within each trajectory cluster.}
    \label{fig:growth}
\end{figure}

Given the potential of two nucleation pathways in the 7-6 system, a natural question to probe is whether the pathways lead to different polymorphs.  We extended the trajectories to the growth phase and examined whether the nucleation trajectory clusters from LeaPP are predictive of the polymorphs in growth.\cite{SarupriaJCTC2025} Figure \ref{fig:growth} shows the average composition profile of trajectories from TC-S1 (Fig. \ref{fig:growth}A-C), TC-S2 (Fig. \ref{fig:growth}D-F), TC-L1 (Fig. \ref{fig:growth}G-I) and TC-L2 (Fig. \ref{fig:growth}J-L). These results reveal that the nucleation trajectory clusters correspond to distinct nucleation pathways that persist in growth, leading to different polymorphic outcomes. In 7--6 system, TC-S1 trajectories remain BCC-dominant, while TC-S2 trajectories maintain a mixed FCC-HCP composition. This indicates that TC-S1 nucleates via a BCC-dominant nucleation pathway, whereas TC-S2 nucleates via a FCC-HCP-dominant nucleation pathway. In contrast, TC-L1 and TC-L2 of the 12--6 system exhibit the same FCC-HCP mixture as TC-S2 during growth, suggesting that they also nucleate via the FCC-HCP-dominant pathway.  

\subsection{Mechanism Behind the 7--6 Polymorphism} \label{sec:7-6poly}

LeaPP offers a powerful framework for uncovering the nuanced mechanisms underlying polymorph selection during nucleation. To understand the origin of this polymorphism, we first sought to identify the point along the nucleation trajectories where LeaPP begins to distinguish between the two pathways. To pinpoint the earliest point where polymorph selection occurs, we analyzed the normalized density of nucleation trajectories along two key coordinates: the fraction of BCC within a given nucleus ($f_\text{BCC}$) and the reaction coordinate (n$_\text{tf}$Q$_\text{6,tf}^\text{cl}$) (see Fig. S15A). Additionally, we overlaid the contour plots of the individual normalized density corresponding to the TC-S1 and TC-S2 trajectories onto this coordinate space. This approach highlights the differences between the two densities and allows us to identify the earliest divergence point, which occurs at n$_\text{tf}$Q$_\text{6,tf}^\text{cl} \sim$ 175.

To contextualize this finding in terms of nucleation progress relative to the end of the trajectory, we projected the normalized density of the nucleation trajectories along n$_\text{tf}$Q$_\text{6,tf}^\text{cl}$ and the committor probability (Fig. S15B). The black horizontal line in this plot marks n$_\text{tf}$Q$_\text{6,tf}^\text{cl}$ = 175, and the earliest corresponding committor value is 0.7.

\begin{figure}[h!]
    \centering
    \includegraphics[width=\linewidth]{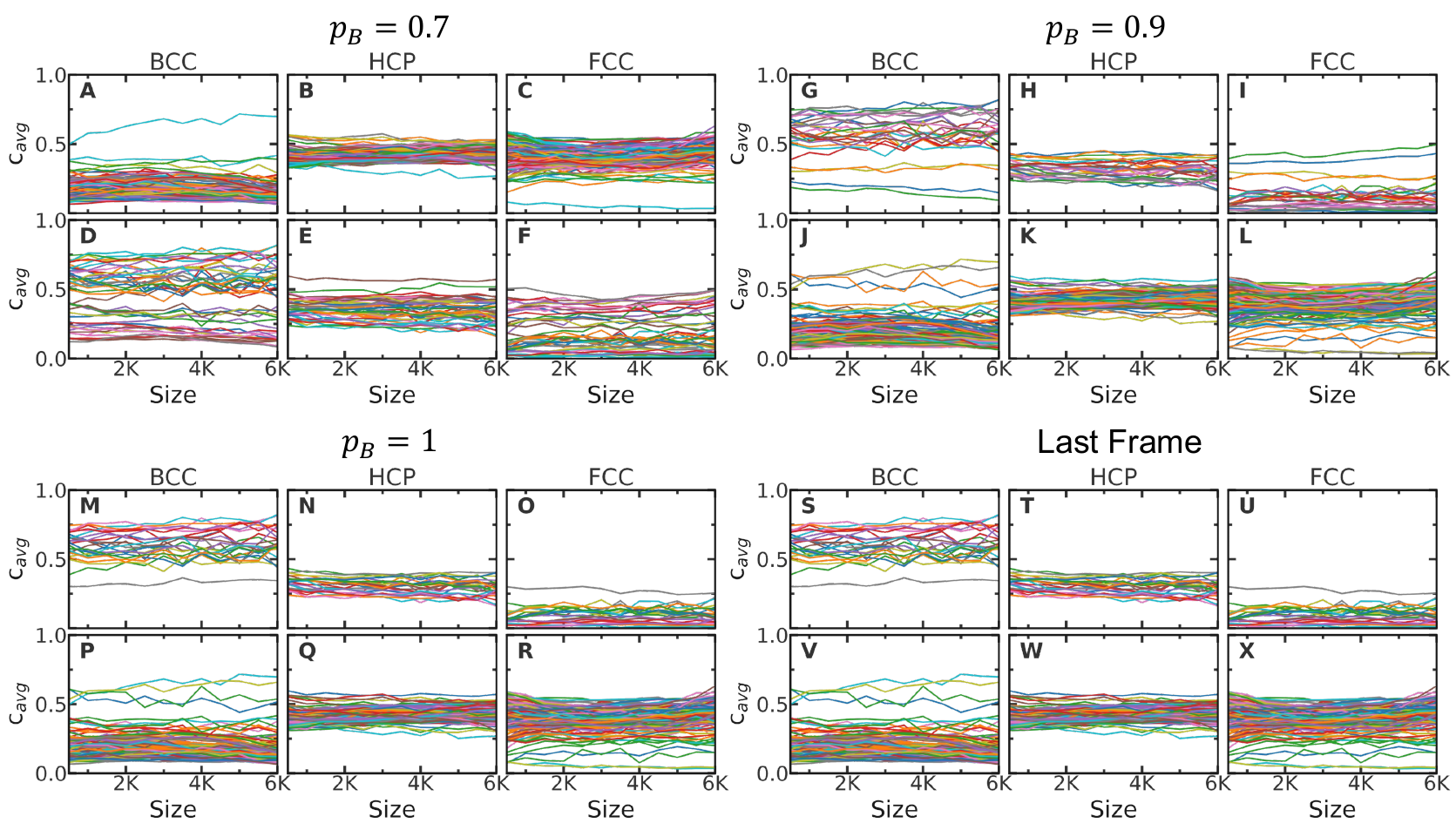}
    \caption{(A-R) Average composition of trajectories in each trajectory cluster based on different stop-frame criteria: the first frame where the committor is (A-F) 0.7, (G-L) 0.9, and (M-R) 1. In (S-X), the stop-frame is the last frame of the trajectory. For each stop-frame criteria, each row represents a trajectory cluster, while each column quantifies the fraction of BCC/HCP/FCC at a given nucleus size. The different stop-frames (committors values of 0.7, 0.9, 1) are chosen based on insights from Fig. S15B, which highlights the regions where the two nucleation pathways are likely to diverge. The average composition profile obtained using the last frame as the stop-frame serves as a reference, representing LeaPP's ground truth.}
    \label{fig:lj-diffstopframe}
\end{figure}

With the earliest committor value for polymorph selection identified, we reapplied Step 3 of LeaPP with one modification. Instead of using the final frame of the trajectory as the stop-frame for the bar graph counts, we used the first frame where the committor criterion is satisfied. As shown in Fig. \ref{fig:lj-diffstopframe}A-F, LeaPP cannot resolve the two distinction pathways when the trajectory information is limited to where the committor is 0.7. Some FCC-HCP-dominant trajectories were clustered in the same trajectory cluster as BCC-dominant trajectories, as shown in Fig.~\ref{fig:lj-diffstopframe}D-F. To improve the resolution, we extended the stop-frames to include committor values of 0.9 (Fig. \ref{fig:lj-diffstopframe}G-L) and 1 (Fig. \ref{fig:lj-diffstopframe}M-R). While the stop-frame at 0.9 shows a significant improvement from 0.7, the classifications in the BCC-dominant pathway (Fig. \ref{fig:lj-diffstopframe}G-I) remain less precise than those at the stop-frame of 1 (Fig. \ref{fig:lj-diffstopframe}M-O). Using the committor value of 1 as the stop-frame achieved the same accuracy as when the last frame of the trajectory is used (Fig. \ref{fig:lj-diffstopframe}S-X).

The results from Fig. \ref{fig:lj-diffstopframe} highlight two important observations. First, the polymorphism of the 7--6 system occurs during post-critical growth. This observation aligns with recent studies~\cite{Molinero2022JPCL, Corti2023JCP}, which show deviation from the classical nucleation picture, where each polymorph nucleates through the critical cluster that is assumed to be structurally identical to its bulk crystal. Second, LeaPP becomes predictive of the polymorphs once it has accumulated information from post-critical growth. These two crucial ideas inspired us to dive further into post-critical growth and understand the driving force behind the 7--6 polymorphism.
\begin{figure}[h!]
    \centering
    \includegraphics[width=\linewidth]{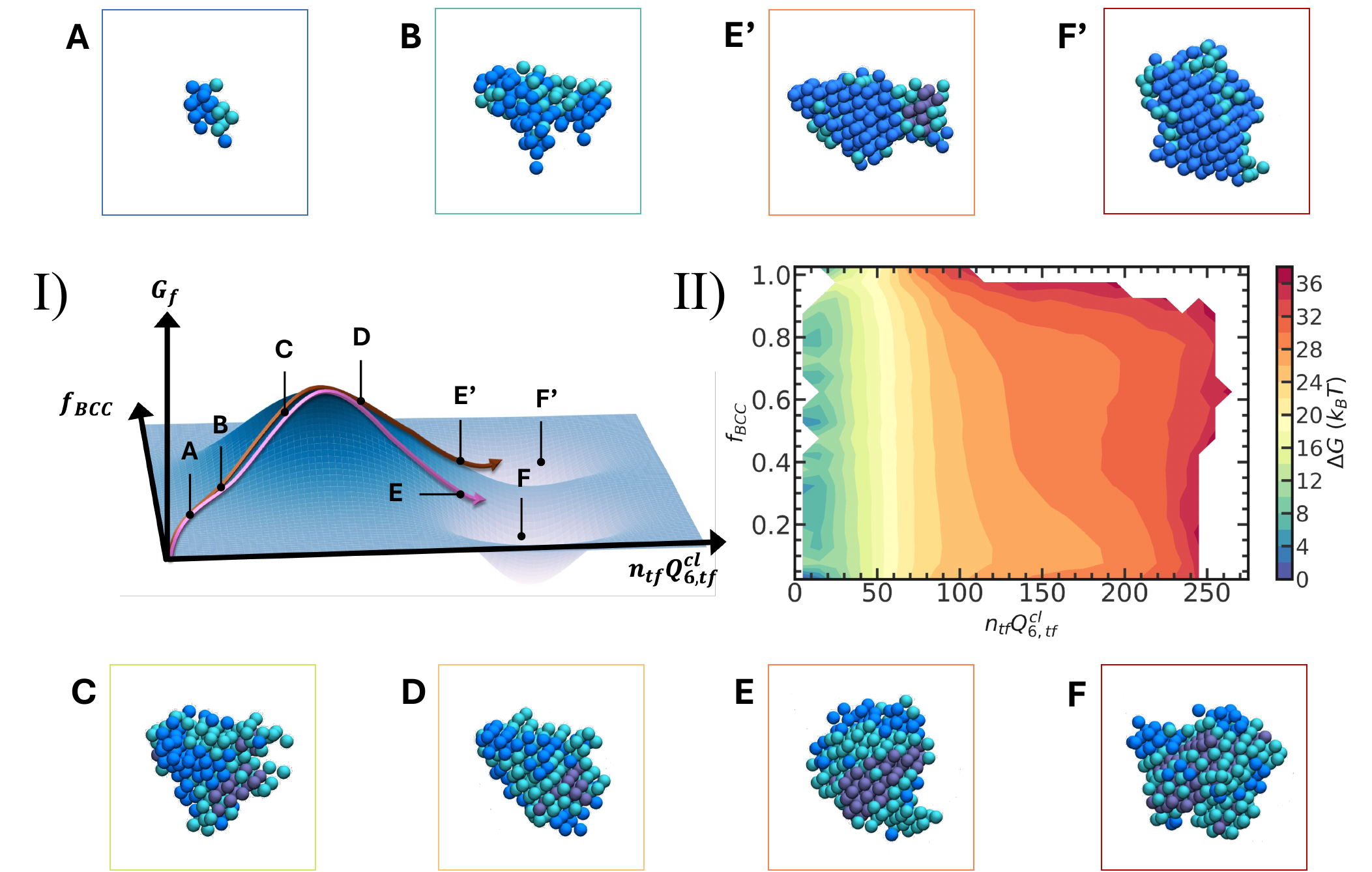}
    \caption{(I) Schematic representation of the proposed free energy landscape along two coordinates: the fraction of BCC in the nucleus ($f_\text{BCC}$) and the reaction coordinate ($n_{tf}Q_{6,tf}^{cl}$). The curve ABCDEF represents the FCC-HCP dominant pathway (TC-S2), and ABCDE'F' represents the BCC dominant pathway (TC-S1). Snapshots along both the pathways are provided in panels A through F, E' and F'. (II) The conditional free energy landscape obtained from simulation data. The conditional free energy is projected onto the same two coordinates as the proposed schematic. The colorbar reflects the free energy relative to the liquid state, expressed in units of $k_{B}$T.}
    \label{fig:lj-polymorph-scheme}
\end{figure}

Based on these findings, we propose the following process for nucleation and growth in the 7-6 system: the nucleation progresses along the reaction coordinate by increasing in size and crystallinity, and after the nuclei grow beyond the critical size, growth follows one of the two pathways -- the BCC dominant pathway (TC-S1) or the FCC-HCP-dominant pathway (TC-S2). The proposed process is schematically depicted in Fig. \ref{fig:lj-polymorph-scheme}(I).

To understand the mechanism governing this post-critical growth, we visualized trajectories from TC-S1 and TC-S2 (see SI Movies S1 and S2). Through this visualization, qualitatively, we observed that the post-critical regime is characterized by the competition between the HCP and BCC phases to capture the core of the nucleus. It is the critical determinant of polymorph selection during post-critical growth. Specifically, a nucleation trajectory becomes FCC-HCP-dominant if the HCP successfully captures the core; enabling FCC to cross-nucleate from HCP as the nucleus continues to grow~\cite{Desgranges_Molecular_2006, Delhommelle:07:JPCB}. Conversely, if the BCC phase captures the core, the trajectory becomes BCC-dominant trajectory. Since cross-nucleation occurs only between phases with similar free energy or stability~\cite{Desgranges_Molecular_2006, Delhommelle:07:JPCB,Desgranges_Controlling_2007}, BCC does not cross-nucleate on HCP or FCC, and vice versa. As such, the phase (HCP/BCC) that captures the core and stabilizes in growth will determine the resulting polymorph.

This phenomenon can also be quantitatively explained through the projection of the conditional free energy along $f_\text{BCC}$ and n$_\text{tf}$Q$_\text{6,tf}^\text{cl}$ (Fig. \ref{fig:lj-polymorph-scheme}(II)). The conditional free energy is obtained from RETIS simulations using path reweighting strategies outlined in Ref. \citenum{Bolhuis:10:JCP} and \citenum{vanErp2016JCTC}. The resulting plot confirms the existence of the two nucleation pathways. It illustrates that as a trajectory progresses away from the basin and approaches the critical size (approximately n$_\text{tf}$Q$_\text{6,tf}^\text{cl} = 125$), no energy barrier separates the two pathways. However, beyond the critical size, a barrier emerges to distinctly separate the two pathways. One pathway leads to a high $f_\text{BCC}$ (BCC-dominant or TC-S1), while the other leads to a low $f_\text{BCC}$ (FCC-HCP-dominant or TC-S2). 

In light of these insights, going back to Fig. \ref{fig:lj-diffstopframe}, we reason that most of the trajectories that reach the committor of 1 have likely resolved the competition between HCP and BCC. This resolution enables LeaPP to accurately capture and learn the evolution of the nucleus composition, such that it can distinctly separate the two nucleation pathways. In contrast, using a committor value of 0.7 or 0.9, where the competition between HCP and BCC may remain unresolved, can result in trajectory clusters that do not correspond to the polymorph observed. For instance, if a nucleus with a relatively high fraction of BCC later grows into an HCP-dominant nucleus, but the stop-frame used for LeaPP analysis is set before the competition is resolved, LeaPP might classify the trajectory as part of the TC-S1 or BCC-dominant pathway.

In summary, these findings highlight two key insights: (i) the competition between HCP and BCC structures within the nucleus core during post-critical growth plays an important role in the 7--6 polymorph selection; and (ii)  LeaPP serves as a valuable tool for determining the point of polymorph selection, and providing a means to assess the sufficiency of information for polymorph prediction.

\section{Conclusions}

We studied the effect of potential softness on crystal nucleation mechanisms and kinetics of Lennard-Jones-like systems under moderate supercooling. Despite differences in interaction potential, we found that both systems exhibit nearly identical nucleation rates at the same pressure and supercooling. Seeding calculations within the CNT framework were used to probe the thermodynamic and kinetic contributions to the nucleation rate. We found that the thermodynamic nucleation barrier controls the nucleation rate, not the kinetic prefactor. The similarity in rates arises from a balance of comparable driving force per unit volume ($\rho_s|\Delta\mu|$) and surface tension ($\gamma$) of the two systems, yielding comparable nucleation barriers.

We analyzed the configurations harvested from RETIS and extracted the reaction coordinate using the MLE method. The OPs that best predicted the progress of nucleation for both systems included a metric of the largest nucleus size and crystallinity. Specifically, n$_\text{tf}$ and Q$_\text{6,tf}^\text{cl}$ combined to represent the reaction coordinate for the 7--6 system and the 12--6 reaction coordinate was represented by n$_\text{ld}$ and f$_\text{cloud}$. The critical nuclei sampled from the 7--6 system displayed a variety of structural compositions, where the critical nuclei were either BCC- or FCC-dominated. Whereas the 12--6 system exclusively formed FCC-dominated critical nuclei. Further analyses from LeaPP show that these two systems nucleate through different nucleation pathways. The 7--6 trajectories can either nucleate through the BCC-dominant pathway or the FCC-HCP-dominant pathway, while the 12--6 trajectories exclusively nucleate through the FCC-HCP-dominant pathway. Furthermore, we also uncovered that the mechanism underlying the polymorphism observed in the 7--6 system is the competition between HCP and BCC to capture the nucleus core at the post-critical stage of the nucleation. Thus, we demonstrate that similar nucleation rates do not necessarily imply identical nucleation mechanisms. In future research, it would be interesting to disentangle the independent effects of the repulsive and attractive components, at constant driving force, on nucleation kinetics and polymorphism. Furthermore, whether the observed independence of nucleation rate from the softness of the potential persists across different supercooling regimes and degrees of softness remains an open and important question.

Overall, this study contributes to a more complete understanding of nucleation under different modifications of intermolecular interactions. The insights gained here contribute to a deeper understanding of polymorph selection and self-assembly mechanisms in simple liquids. Notably, the ability to pinpoint the moment of polymorph selection and uncover the underlying mechanisms offers an unprecedented level of resolution in the study of nucleation. Together, these findings offer fundamental guidance for future research in crystallization and phase transitions.  

\begin{acknowledgement}

This material is based on work supported by NSF CBET No. 1653352. This work was supported by the U.S. Department of Energy, Office of Science, Office of Basic Energy Sciences, under Award No. DE-SC0015448 and DE-SC0025496. S.W.H. and S.S. acknowledge the Data Science Initiative at the University of Minnesota College of Science and Engineering for funding through the ADC Graduate Fellowship. The authors acknowledge the University of Minnesota Start-up funds and the Minnesota Supercomputing Institute at the University of Minnesota for providing resources that contributed to the reported results.

\end{acknowledgement}

\section{Data Availability Statement}
The full dataset supporting the findings of this study is available from the corresponding author upon reasonable request.

\suppinfo
The supporting information contains supporting figures and tables, as well as implementation and calculation details for various analyses mentioned in the main text.

\providecommand{\latin}[1]{#1}
\makeatletter
\providecommand{\doi}
  {\begingroup\let\do\@makeother\dospecials
  \catcode`\{=1 \catcode`\}=2 \doi@aux}
\providecommand{\doi@aux}[1]{\endgroup\texttt{#1}}
\makeatother
\providecommand*\mcitethebibliography{\thebibliography}
\csname @ifundefined\endcsname{endmcitethebibliography}
  {\let\endmcitethebibliography\endthebibliography}{}

\end{document}


\maketitle

\section{Solid-liquid equilibrium}

Solid--liquid phase diagrams for the two potentials are required to
select appropriate conditions to study crystal nucleation. To our
knowledge, no phase diagram is available for the 7--6 potential
utilized in this work. Though phase diagrams are available for the
12--6 potential, our use of a force-switch and relatively long
interaction cut-off distance results in slightly different melting
points for our system.\cite{Ahmed_Effect_2010} Melting temperatures
were thus calculated for a range of pressures from $p=1$ to $p=20$
with direct coexistence between the liquid and FCC phases.\cite{Fernndez_The_2006}

\begin{figure}[htbp]
\begin{center}
\includegraphics[width=0.8\linewidth]{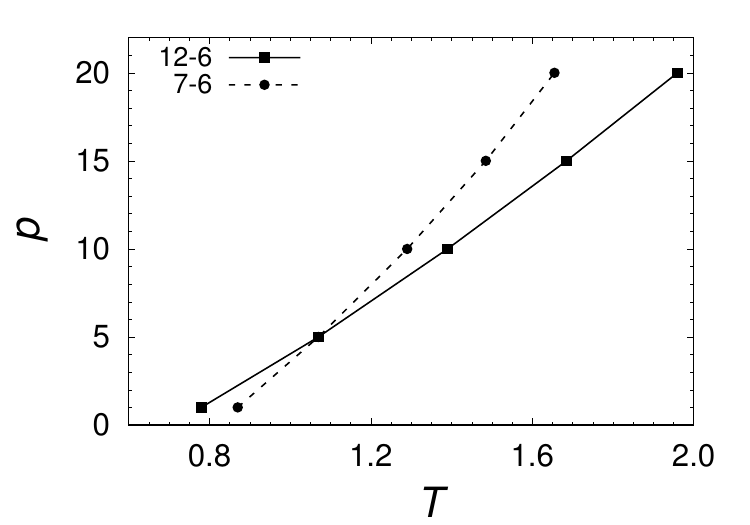}
\caption{Temperature-pressure phase diagram of the $n$-6 potentials under study. Error bars are smaller than the symbols.}
\label{fig:phaseDiagram}
\end{center}
\end{figure}

The direct coexistence method takes advantage of the fact that if two
phases are in contact at an interface of constant area, the phase with
lower free energy will grow. To identify the melting point at any one
pressure, straightforward simulations are performed with the two
phases in contact across a range of temperatures. At the melting
point, neither phase will grow. 

The resulting melting points are plotted in
Fig.~\ref{fig:phaseDiagram}. There is a crossover point between the
two melting curves at $p=5$. Below $p=5$, the 7--6 potential has the
higher melting temperature ($T_{\text{m}}$) of the two potentials,
whereas above $p=5$ the 12--6 potential has the higher
$T_{\text{m}}$. We chose to study nucleation at $p=5$, where
$T_{\text{m}}$ is equal for the two potentials. This choice has the
benefit that the same degree of supercooling is the same temperature
for both potentials, allowing for a more direct comparison.

\section{Direct coexistence simulations}

\begin{figure}[h!]
\begin{center}
\includegraphics[width=0.8\linewidth]{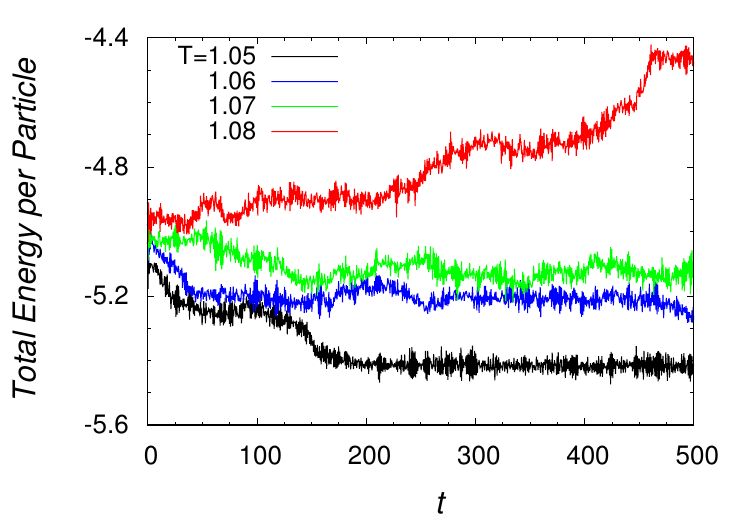}
\caption{Evolution of total energy per particle in the 12-6 direct coexistence simulations at a pressure of 5.}
\label{fig:ener-126}
\end{center}
\end{figure}

\begin{figure}[h!]
\begin{center}
\includegraphics[width=0.8\linewidth]{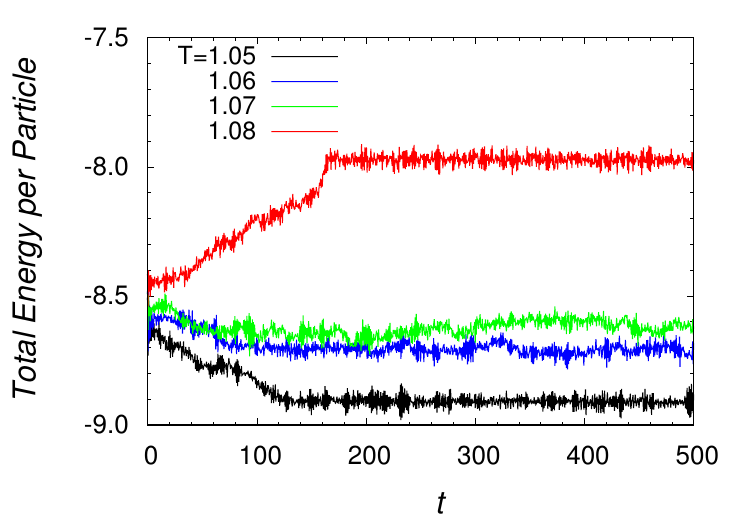}
\caption{Evolution of total energy per particle in the 7-6 direct coexistence simulations at a pressure of 5.}
\label{fig:ener-76}
\end{center}
\end{figure}

At each pressure of interest, direct coexistence simulations were run across a range of temperatures. The initial configuration was generated by contacting a liquid configuration of 4000 particles with the (100) plane of an equilibrated FCC crystal containing 3872 particles. Energy minimization was performed on the liquid particles to remove high-energy overlaps at the interfaces, followed by a 500-$\tau$ $Np_zT$ simulation where box fluctuations were only allowed in the z-direction. At temperatures above the melting temperature, the solid region melts and the total energy increases as the thermostat resupplies kinetic energy to the system. The opposite occurs below the melting temperature, causing a decrease in the energy. By exploiting this behavior, the melting temperature is determined by finding the lowest temperature at which the solid melts and the highest at which it grows. The two are averaged to obtain the melting temperature with an error equal to half the difference between the temperatures. Example energy plots from simulations are shown in Fig. \ref{fig:ener-126} and Fig. \ref{fig:ener-76}.

\clearpage

\section{Local Crossing Histogram}

\begin{figure}[h!]
\begin{center}
\includegraphics[width=\linewidth]{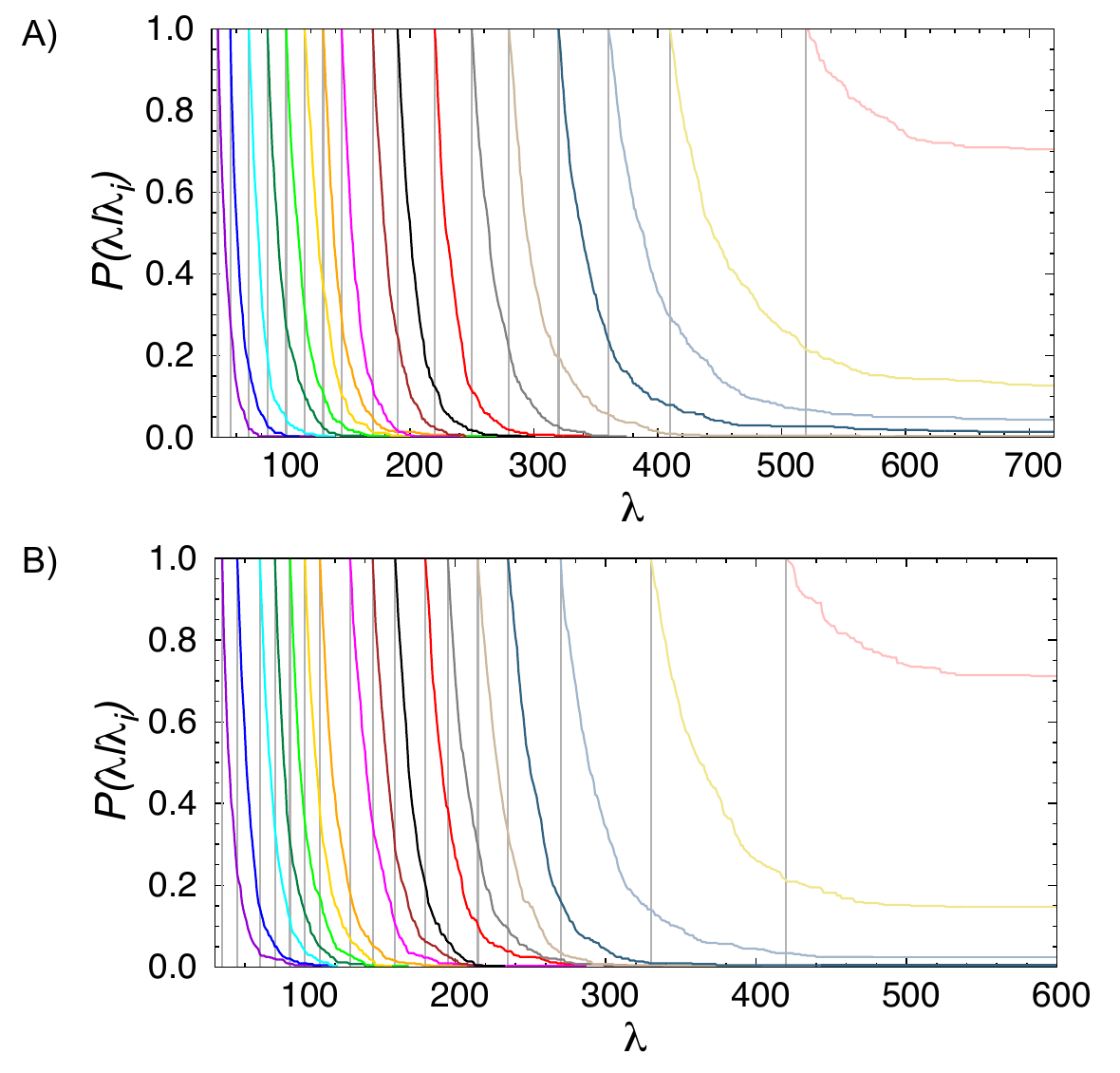}
\caption{\textbf{Local crossing probability histogram for each interface ensemble.} This crossing histogram is constructed from every 10th path in 4400 MC moves such that the crossing histogram only reflects decorrelated paths sampling for the ensemble. Each vertical line represents each interface ensemble $\lambda_i$. Each colored line represents the crossing probability of paths from ensemble $\lambda_i$ to the next higher $\lambda$ value (i.e., $P(\lambda|\lambda_i)$). The crossing probability histogram for the 7--6 system is shown in panel (A), and the crossing probability histogram for the 12--6 potential is shown in panel (B).\cite{SarupriaJCP2022}}
\label{fig:crossingprob}
\end{center}
\end{figure}
\clearpage

\section{Seeding}

\subsection{Sanz's Seeding Procedure}
The surface free energies are calculated using the seeding method~\cite{Espinosa_The_2015,Espinosa_Seeding_2016}. This approach relies on the validity of CNT, but it allows the calculation of the surface free energy away from coexistence. The procedure is outlined below:
\begin{enumerate}
\item Generate a liquid configuration of 45,834 randomly placed particles.
\item Equilibrate the liquid for 500 $\tau$ with Berendsen thermostat and barostat at $T = 0.85T_m$ ($T_m=1.07\epsilon/k_B$) and $p = 5\epsilon/\sigma^3$.
\item Equilibrate the liquid for 1000 $\tau$ with Nos\'{e}-Hoover thermostat and Parrinello-Rahman barostat at the same temperature and pressure as in step 2.
\item Cut a spherical seed from a perfect FCC configuration with 1.55$\sigma$ lattice spacing (1.578$\sigma$ at $T_m$).
\item Insert the seed into the final frame from the liquid equilibration simulation, removing any liquid particles within 1$\sigma$ of the seed.
\item Freeze the positions of the seed particles and equilibrate the liquid particles for 7$\tau$ with Berendsen thermostat and barostat at $T=0.85T_m$ and $p = 5\epsilon/\sigma^3$.
\item Equilibrate the system with no frozen particles for 7$\tau$ with Berendsen thermostat and barostat at the same temperature and pressure as in step 6.
\item For a range of temperatures expected to contain the critical temperature, launch one 150-$\tau$ trajectory at each temperature, in increments of $8.314\times10^{-3}\epsilon/k_B$ using the GROMACS V-rescale thermostat and Parrinello-Rahman barostat.
\item At the temperature where the nucleus size appears roughly constant with time, select temperatures in increments of 0.0017$\epsilon/k_B$ and launch eight trajectories from the same seed configuration with randomized velocities for each temperature. The temperature where this nucleus is critical is where between 3/8 and 5/8 of the trajectories show growth. We denote this temperature as $T^*$.
\item Calculate the FCC density at $T^*$ by running a 1000-$\tau$ NpT simulation of the FCC crystal with Nos\'{e}-Hoover thermostat and Parrinello-Rahman barostat after a 500-$\tau$ equilibration.
\item Calculate the FCC-liquid chemical potential difference at $T^*$ with thermodynamic integration of the Gibbs-Helmholtz equation.\cite{Sanz2021PCCP}
\item Calculate the surface free energy using $\gamma = (\frac{3}{32\pi})^{1/3} n_c^{1/3} \rho_\text{s}^{2/3} |\Delta \mu|$.
\end{enumerate}

The calculated surface free energy depends on the size of the critical nucleus. Consequently, this method is sensitive to the definition of crystallinity used. In this work, $q_{6,avg}$\cite{Lechner_Accurate_2008} distributions are calculated for the bulk FCC and bulk liquid phases at each temperature a critical nucleus is obtained using a neighbor cut-off distance equal to the first minimum of the FCC radial distribution (12-6: 1.36$\sigma$, 7-6: 1.28$\sigma$). The $q_{6,avg}$ cut-off for solid classification is then chosen to be the midpoint between the values where the FCC and liquid $q_{6,avg}$ distributions become zero. 

\begin{table}[h]
\linespread{0.9}\selectfont\centering
\caption{12-6 seeding results at lower supercooling}
\begin{center}
\setlength{\tabcolsep}{6pt}

\begin{tabular}{@{} c  c  c  c  c @{}} \toprule
$n_{\text{c}}$ & $T_{\text{c}}$ & $\rho_{\text{s}}$ & $|\Delta\mu|$ & $\gamma(T_{\text{c}})$\\
\midrule
816  &  0.923  &  1.042  &  0.184  &  0.547 \\
2312  &  0.964  &  1.036  &  0.133  &   0.556 \\
3635  &  0.979  &  1.033  &  0.115  &  0.561  \\
6233  &  0.994  &  1.031  &  0.0968  &   0.564 \\
\bottomrule

\label{tab:12-6-gamma}
\end{tabular} 
\end{center}  
\end{table} 

\begin{table}[h]
\linespread{0.9}\selectfont\centering
\caption{7-6 seeding results at lower supercooling}
\begin{center}
\setlength{\tabcolsep}{6pt}

\begin{tabular}{@{} c  c  c  c  c@{}} \toprule
$n_{\text{c}}$ & $T_{\text{c}}$ & $\rho_{\text{s}}$ & $|\Delta\mu|$ & $\gamma(T_{\text{c}})$\\
\midrule
882  &  0.906  &  1.243  &  0.173  &  0.596 \\
1587  &  0.932  &  1.240  &  0.146  &   0.610 \\
4499  &  0.971  &  1.234  &  0.105  &  0.619  \\
5796  &  0.979  &  1.232  &  0.0972  &   0.623 \\
\bottomrule
\label{tab:7-6-gamma}
\end{tabular} 
\end{center}  
\end{table} 

\begin{figure}[h!]
    \includegraphics[width=0.7\textwidth]{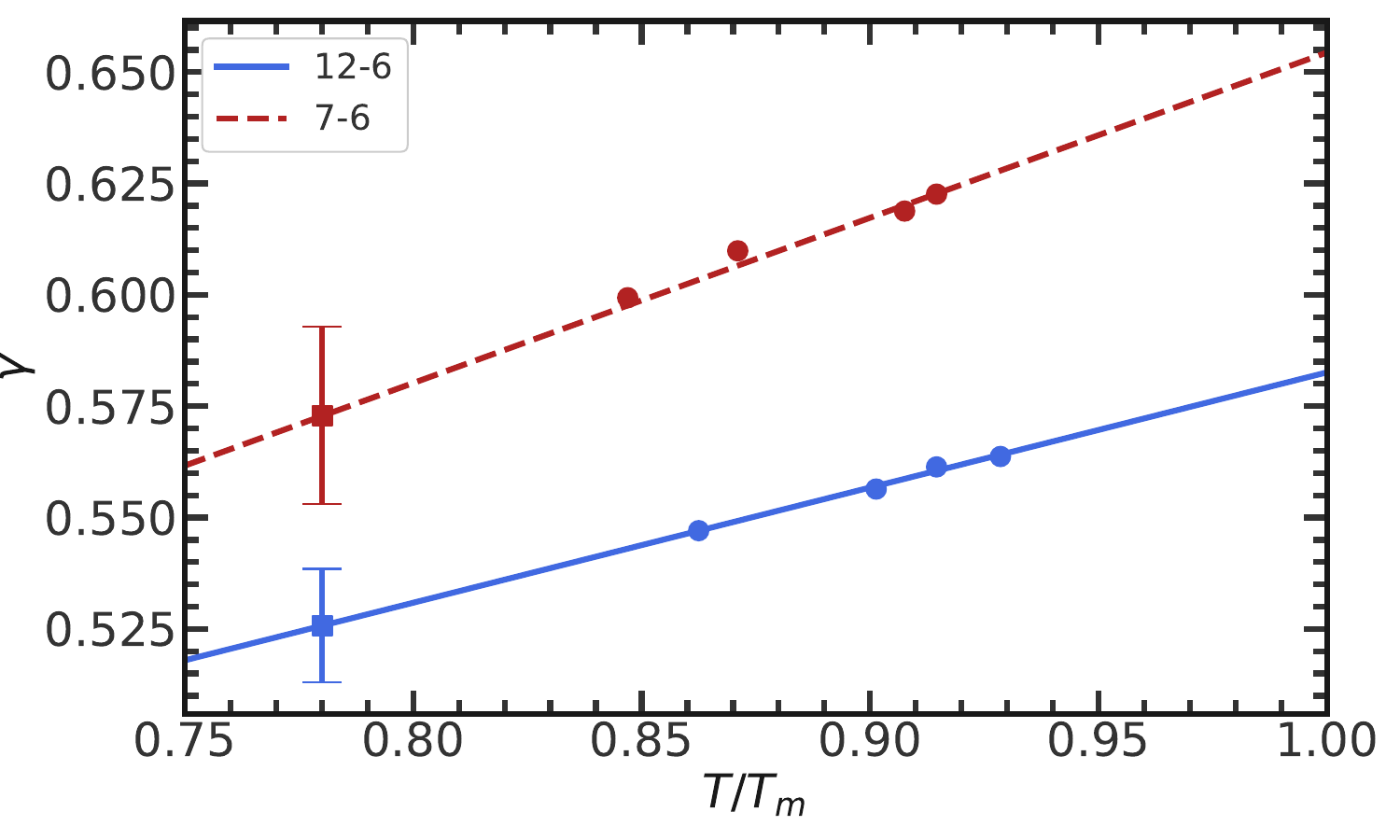}
    \caption{Surface free energies obtained from seeding simulations (round points) in Table~\ref{tab:12-6-gamma} and~\ref{tab:7-6-gamma} and the extrapolations to 0.78Tm (square points). Error bars were calculated for 95\% confidence level.}
    \label{fig:gammavtemp}
\end{figure}
\newpage

\subsection{Calculation of attachment rate (D)}
The attachment rate was calculated at the same conditions as RETIS simulations with the procedure described in
Ref. \citenum{Zimmermann_NaCl_2018}. Spherical seeds with sizes
ranging from approximately 50--500 particles were obtained from bulk FCC crystals which had been
equilibrated at the conditions of interest. The seeds were inserted
into equilibrated bulk liquid configurations and overlapping particles
were removed. The seed--liquid system was carefully equilibrated to
relax the solid--liquid interface. Below describes the complete details of the
equilibration procedure:

\begin{enumerate}
\item Equilibrate liquid configuration for 500$\tau$ at RETIS temperature and collect ten configurations from a 1000-$\tau$ simulation launched from the equilibrated configuration.
\item Repeat for a bulk FCC system, but cut spherical seeds of different sizes from ten configurations, spaced 100$\tau$ apart.
\item Combine the ten seeds with the ten liquid boxes, producing ten seeded configurations.
\item Equilibrate seeded configurations in three blocks with three steps per block:
\begin{enumerate}
\item Seed frozen, liquid moving: 0.1$\tau$ with $\Delta t = 10^{-4}\tau$, 0.125$\tau$ with $\Delta t = 2.5\times10^{-4}\tau$, and 8$\tau$ with $\Delta t = 10^{-3}\tau$
\item Liquid frozen, seed moving: same as (a), but in NVT ensemble because GROMACS does not scale the coordinates of frozen atoms, which are the majority of the configuration
\item All particles moving: same as (a)
\end{enumerate}
\end{enumerate}

\begin{figure}[h!]
\begin{center}
\includegraphics[width=0.49\textwidth]{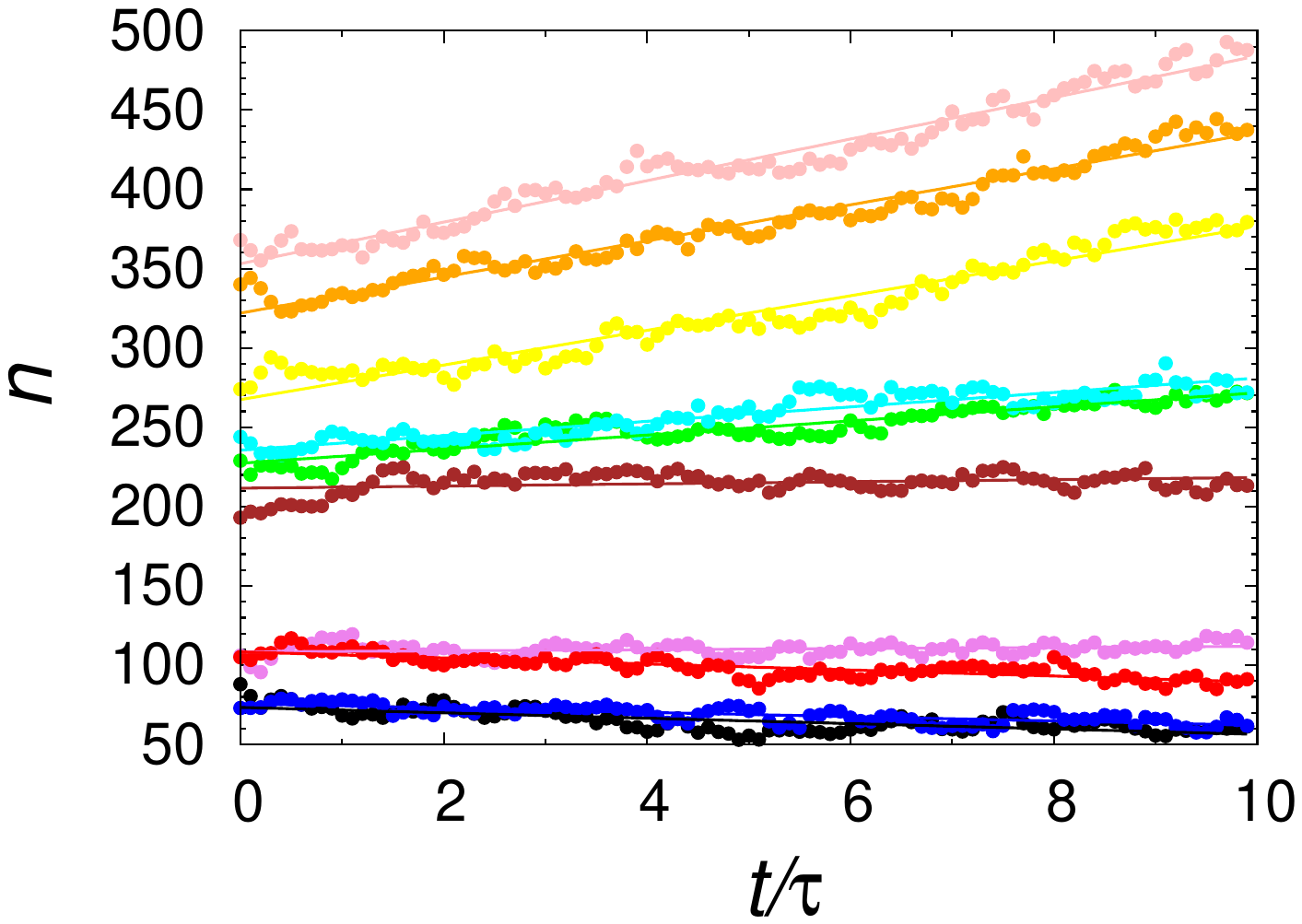}
\includegraphics[width=0.49\textwidth]{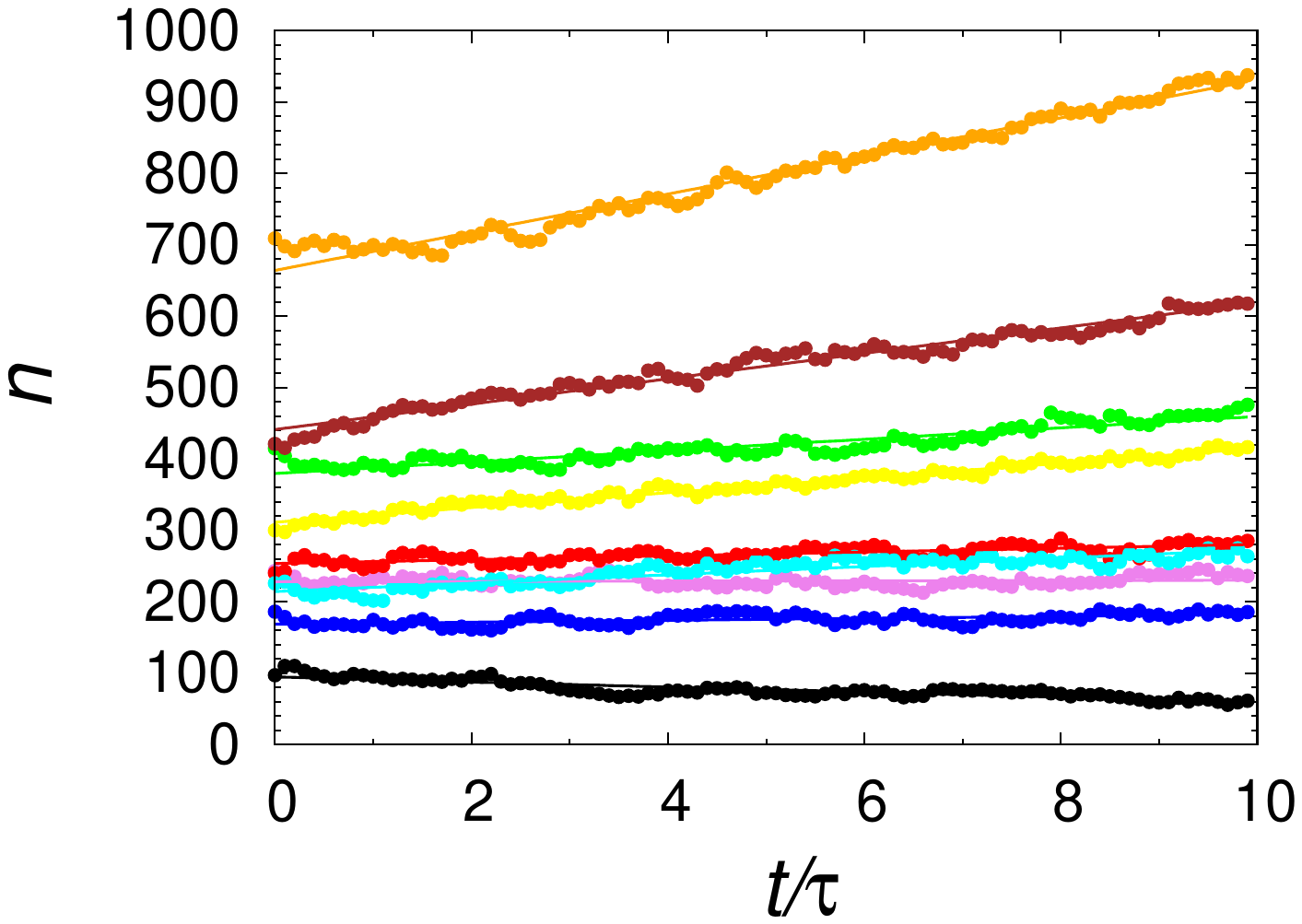}
\caption{Linear fits to cluster size versus time, each averaged over ten trajectories. Left column: 12--6, right column: 7--6.}
\label{fig:avgClusterSize}
\end{center}
\end{figure}

\begin{figure}[h!]
\begin{center}
\includegraphics[width=0.49\textwidth]{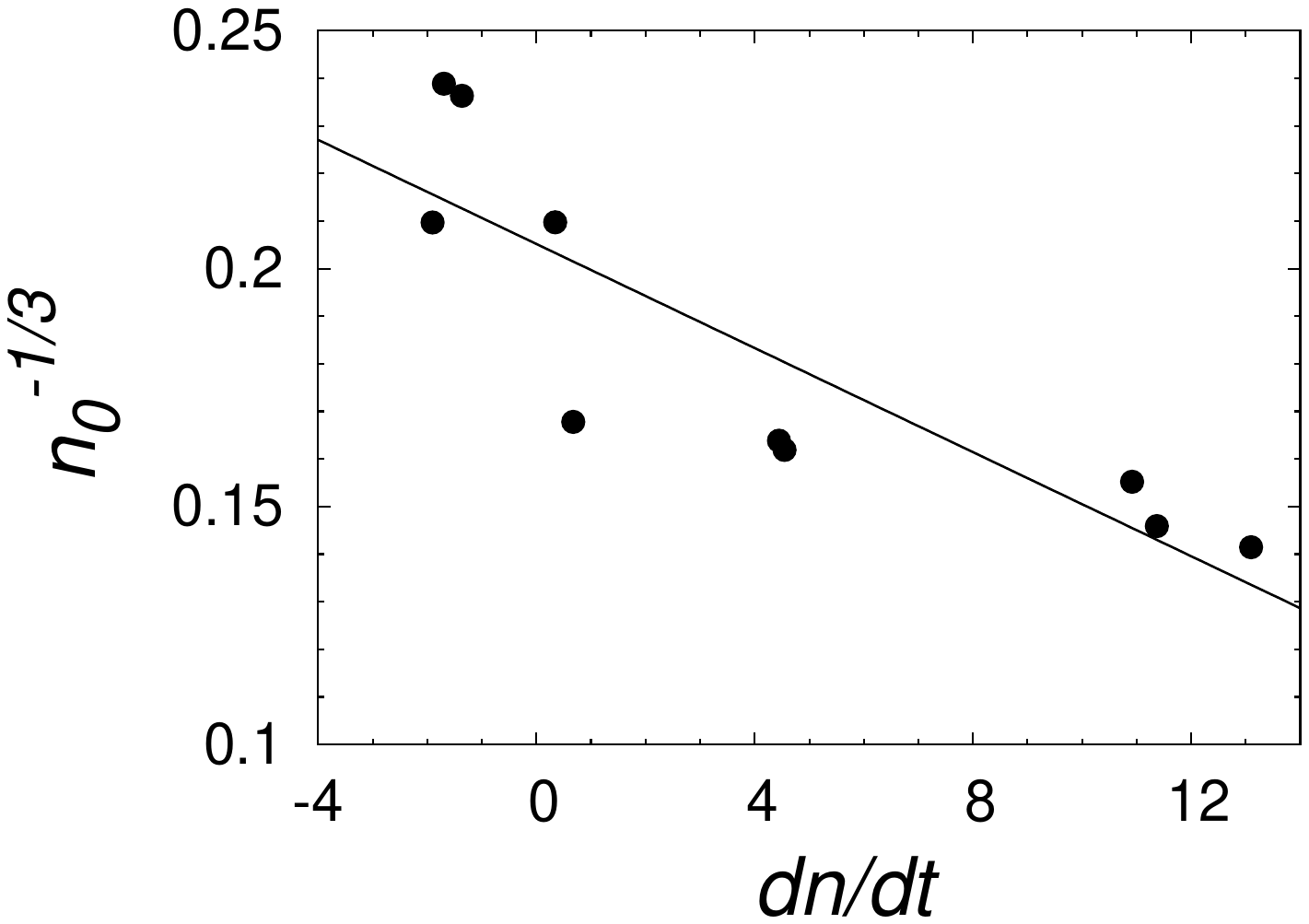}
\includegraphics[width=0.49\textwidth]{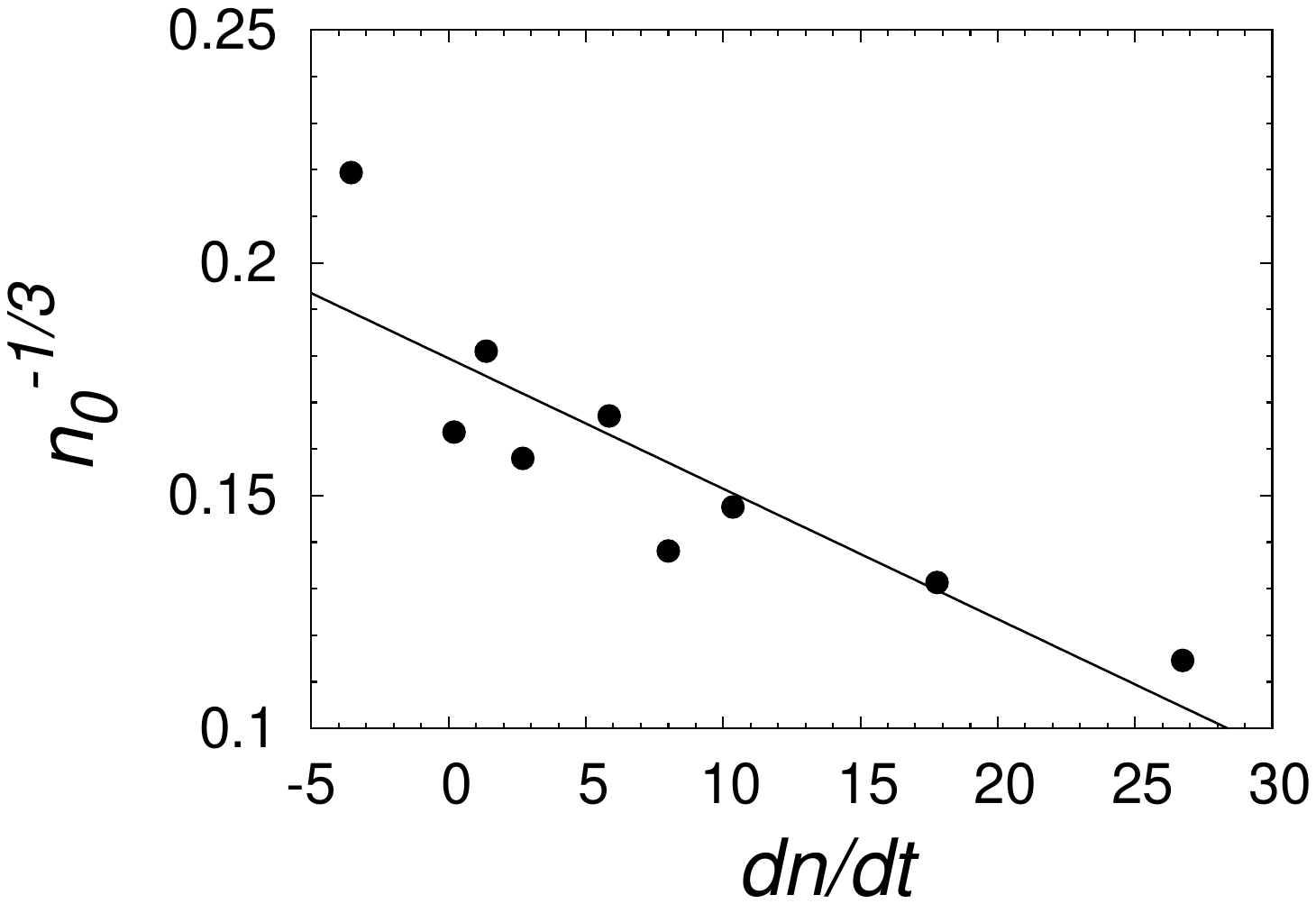}
\caption{Linear fits to $n_0^{-1/3}$ versus $dn/dt$ calculated from the linear fits in the previous figure, where $n_0$ is the intercept obtained from the linear fit. Left column: 12--6, right column: 7--6.}
\label{fig:dndt}
\end{center}
\end{figure}

For each equilibrated seed--liquid system, ten trajectories were initiated with different momenta randomly drawn in accordance with the Maxwell--Boltzmann distribution. The size of each seed was monitored with time and averaged across the ten trajectories to determine the average nucleus size vs. time behavior for each initial seed size. The evolution of the seed sizes were fit to lines to calculate $dn/dt$ (slope) and initial nucleus size $n_0$ (intercept) for each seed size
(Fig. \ref{fig:avgClusterSize}). The relationship between
$dn/dt$ and $n_0$ across a range of seed sizes were used with
Eq. \ref{eq:seed1-final}: 
\begin{equation}
n_0^{-1/3}=\frac{3}{2}\frac{1}{\phi \gamma} \Bigg[ |\Delta \mu| - \frac{k_\text{B}T}{D}\frac{dn}{dt}\Bigg].
\label{eq:seed1-final}
\end{equation}
to estimate $D$ and $\phi \gamma$
(Fig. \ref{fig:dndt}). Though $\phi \gamma$ is not required to
estimate the nucleation rate, $J$, from
Eq. 6, knowledge of $\phi \gamma$ is useful
in understanding the interplay between the driving forces for nucleation (see Eq. 3). $D$ is expressed in units of $n^2/\tau$. Note that $D$ is assumed to be independent of $n$ in the range of nucleus sizes tested.

\newpage

\section{Order parameters} \label{sec:ops}

In this work, order parameters (OP) were used to perform both RETIS and further analyze the nucleation mechanism once sampling was
completed. Following common practice, the RETIS OP was the size of the largest crystalline nucleus in the system. To distinguish
from other nucleus-size-based metrics, we denote this sampling OP as $n_\text{tf}$\cite{ten_Wolde_Numerical_1996,tenWolde:05:PRL,SarupriaJCTC2025}. The largest nucleus size metric based on the protocol from Ref. \citenum{Lechner_Accurate_2008} is denoted as $n_\text{ld}$.

\subsection{$n_\text{tf}$}  \label{sec:ntf}
The detailed procedure for calculating this largest nucleus size is described in Ref.~\citenum{SarupriaJCTC2025}. Here we briefly describe the procedure. First, we need to identify solid particles that exist within the configuration. We calculate the complex vector, $q_{6m}$, for each particle i. 
\begin{equation}
q_{6m}(i)=\frac{1}{N_n(i)}\sum_{j=1}^{N_n(i)} Y_{6m}(\textbf{r}_{ij}),
\label{eq:q6m}
\end{equation}
where $Y_{6m}$ are the $l=6$ spherical harmonics, $\textbf{r}_{ij}$ is the unit vector between particles $i$ and $j$, and $N_n$ is the number of neighbors within the first neighbor shell ($r_\text{cut}$) of particle $i$ according to the liquid RDF. The resulting complex vector is 13-dimensional since the $l=6$ spherical harmonics span integers from $m=-6$ to $m=6$. Then the normalized $q_{6m}$ dot product of each neighbor pair,
\begin{equation}
d_{ij}=\frac{\sum_{m=-6}^{m=6} q_{6m}(i)q_{6m}(j)^{*}}{\big(\sum_{m=-6}^{m=6} |q_{6m}(i)|^2\big)^{1/2}\big(\sum_{m=-6}^{m=6} |q_{6m}(j)|^2\big)^{1/2}}
\end{equation}
is calculated to determine if two particles are structurally correlated or ``connected". Two particles $i$ and $j$ are considered connected if $d_{ij} > d_\text{cut}$. A particle $i$ is then considered a solid particle if it has more than $n_\text{connect,cut}$ connections.

Once all solid particles in the system are identified with the above procedure, a graph is constructed with solid particles as nodes. Edges are placed between nodes that are neighbors within $r_\text{cut}$. The largest cluster of solid particles is the largest connected component of the graph. The neighbor cut-off radius, $r_\text{cut} = 1.50\sigma$ for the 12--6 potential, and $r_\text{cut} = 1.42\sigma$ for 7--6. $d_{ij} = 0.5$ and $n_\text{connect,cut} = 9$ are used for both potentials.

\subsection{$n_\text{ld}$}

$n_{\text{ld}}$, like $n_{\text{tf}}$, is another metric for the size of the largest solid nucleus. However, each particle $i$ is classified as solid or not based on the value of the neighbor--averaged bond order parameter $\overline{q_6}(i)$ as described by Lechner and Dellago:\cite{Lechner_Accurate_2008}

\begin{equation}
\overline{q_6}(i)=\bigg(\frac{4\pi}{13}\sum_{m=-6}^{m=6} |\overline{q_{6m}}(i)|^{2}\bigg)^{1/2}
\end{equation}
where

\begin{equation}
\overline{q_{6m}}(i)=\frac{1}{N_n(i)+1}\sum_{j=1}^{N_n(i)+1} q_{6m}(i).
\end{equation}
The sum is over the $N_n$ neighbors of particle $i$ and itself. $q_{6m}(i)$ is calculated as

\begin{equation}
q_{6m}(i)=\frac{1}{N_n(i)}\sum_{j=1}^{N_n(i)} Y_{6m}(\textbf{r}_{ij}),
\label{eq:q6m}
\end{equation}
Note that $\overline{q_{6m}}(i)$ is also a 13-dimensional complex vector. The distribution of $\overline{q_6}(i)$ is calculated for the pure liquid and crystal phases. Cut-off values of $\overline{q_6}(i)$ are selected to distinguish the solid and liquid phases based on the overlap of their distributions.
Once solid particles have been identified based upon their $\overline{q_6}(i)$ values, the largest nucleus of solid particles is
identified with the same approach as described in Sec. \ref{sec:ntf}. $n_\text{ld}$ is the size of the largest nucleus of solid particles. The neighbor cutoff distance, $r_\text{cut}$ was selected as the first minimum of the FCC RDF, which is 1.36$\sigma$ for 12--6 and 1.28$\sigma$ for 7--6. Particles with $\overline{q_6}(i) > 0.36$ were considered solid for both potentials.

\subsection{$n_\text{BCC}$, $n_\text{HCP}$, $n_\text{FCC}$}

$\overline{q_6}(i)$ can differentiate between liquid and crystal particles, but another bond OP is required to distinguish between BCC, HCP, and FCC structures. For particles that are identified as belonging to the largest nucleus identified by $n_{\text{ld}}$, the local structure (BCC, HCP, FCC) of each particle is determined by $\overline{q_4}(i)$. Analogously to $\overline{q_6}(i)$, $\overline{q_4}(i)$ is calculated as :

\begin{equation}
\overline{q_4}(i)=\bigg(\frac{4\pi}{9}\sum_{m=-4}^{m=4} |\overline{q_{4m}}(i)|^{2}\bigg)^{1/2},
\end{equation}
where

\begin{equation}
\overline{q_{4m}}(i)=\frac{1}{N_n(i)+1}\sum_{j=1}^{N_n(i)+1} q_{4m}(i),
\end{equation}
where

\begin{equation}
q_{4m}(i)=\frac{1}{N_n(i)}\sum_{j=1}^{N_n(i)} Y_{4m}(\textbf{r}_{ij}).
\end{equation}
$Y_{4m}$ are the $l=4$ spherical harmonics, and $q_{4m}(i)$ is a 9-dimensional complex vector. Distributions of $\overline{q_4}(i)$ are calculated for the BCC, HCP, and FCC phases. Cut-off values of $\overline{q_4}(i)$ for each phase is selected to be the lower and upper bounds of where the distribution intersects the distribution of another phase. Particles are classified as BCC if $\overline{q_4}(i) < 0.075$, HCP if $0.075 \leq \overline{q_4}(i) < 0.12$, and FCC if $\overline{q_4}(i) \geq 0.12$. Once each particle is classified, the number of BCC, HCP, and FCC particles in each nucleus is calculated.

\subsection{$\kappa_\text{tf}$, $\kappa_\text{ld}$} \label{sec:asphericity}

The nucleus asphericity, $\kappa$, was also calculated. $\kappa$
quantifies the deviation of a collection of points from a spherical
shape, with zero indicating a perfectly spherical nucleus and one
indicating a linear chain. The asphericity is calculated by

\begin{equation}
\kappa=\frac{3}{2}\frac{\alpha_1^4+\alpha_2^4+\alpha_3^4}{(\alpha_1^2+\alpha_2^2+\alpha_3^2)^2}-\frac{1}{2},
\end{equation}
where $\alpha_1$, $\alpha_2$ and $\alpha_3$ are the eigenvalues of the
gyration tensor. Asphericity was calculated for nuclei identified by
both $n_{\text{tf}}$ and $n_{\text{ld}}$. 

\subsection{$Q_{6,\text{tf}}^\text{cl}$, $Q_{6,\text{ld}}^\text{cl}$} \label{sec:crystallinity}
$Q_6^{\text{cl}}$ is a measure of the degree of crystallinity of a nucleus and is calculated by averaging the $Y_{6m}$ spherical harmonics over all bonds between solid neighbors in the nucleus.~\cite{Moroni_Interplay_2005, SarupriaJCTC2025} Details of this calculation on the nucleus is described in Ref.~\citenum{SarupriaJCTC2025}.

\subsection{f$_\text{cloud}$}
As described in Ref.~\citenum{Bolhuis2011PRL}, cloud particles are surface particles that are intermediate between liquid and bulk solid in their structure. By definition, the number of cloud particles within a solid nucleus is calculated as the difference between $n_\text{tf}$ and $n_\text{ld}$.\cite{Bolhuis2011PRL} Thus, we calculate the fraction of cloud particles within a solid nucleus as:
\begin{equation}
    f_\text{cloud} = (n_{tf}-n_{ld})/n_{tf}
\end{equation}

\section{Maximum Likelihood Estimation (MLE)} \label{sec:MLE_imp}

To implement the Peters and Trout MLE method\cite{Peters_Obtaining_2006}, all the different OPs mentioned above in Section \ref{sec:ops} were calculated for all configurations sampled for both systems. After the OPs were calculated, MLE was performed on the three different OP combination models (Table \ref{tab:mle-op-1} and \ref{tab:mle-op-2}). The first model is 
\begin{equation}
    r_1(\textbf{q})=\alpha_1\times \textbf{q} + \alpha_0
\label{eq:rc_1}
\end{equation}
where each trial reaction coordinate (RC) ($r_i$) uses one OP ($q_i$) from the nine calculated. The second model is 
\begin{equation}
    r_2(\textbf{q}_i,\textbf{q}_j)=\alpha_2 \times \textbf{q}_j + \alpha_1 \times \textbf{q}_i +\alpha_0
\label{eq:rc_2}
\end{equation}
where each $r_{ij}$ uses a linear combination of a pair of different OPs ($q_i$, $q_j$). Lastly, the third model is 
\begin{equation}
    r_3(\textbf{q}_i,\textbf{q}_j)=\alpha_1\times \textbf{q}_i \times \textbf{q}_j + \alpha_0 = \alpha_1\times \textbf{q}_{ij} + \alpha_0
\label{eq:rc_3}
\end{equation}
where each $r_{ij}$ uses a nonlinear combination of a pair of different OPs ($q_i$, $q_j$) to create a new OP, $q_{ij}$. MLE was performed for one model at a time to determine the best trial RC from each model (Table \ref{tab:mle-op-1} and \ref{tab:mle-op-2}).

\subsection{MLE result without the inclusion of $f_{cloud}$}

\begin{table}[h]
\caption{Top three ranked OPs for each choice of the trial RC ($\textbf{r}$) for the 7--6 system.}
\centering
      
\begin{tabular}{@{} l c c c c c c @{}} \toprule

& \multicolumn{2}{c}{$r_i=\alpha_1\times q_i+\alpha_0$} & \multicolumn{2}{c}{$r_{ij}=\alpha_2\times q_j +\alpha_1\times q_i+\alpha_0$} & \multicolumn{2}{c}{$r_{ij}=\alpha_1\times q_{ij}+\alpha_0$} \\
\cmidrule{1-3}
\cmidrule{4-5}
\cmidrule{6-7}
Rank & OP & -LL score  & OP & -LL score & OP & -LL score\\

\midrule
1 & $n_\text{tf}$&$7.050\times10^{-5}$&  $n_\text{tf}$ + $Q_\text{6, ld}^\text{cl}$&$6.771\times10^{-5}$& $n_\text{tf}Q_\text{6, tf}^\text{cl}$&$6.956 \times 10^{-5}$\\

2 & $n_\text{ld}$&$7.780\times10^{-5}$&  $n_\text{tf}$ + $\kappa_\text{tf}$&$6.832\times10^{-5}$& $n_\text{ld}Q_\text{6, ld}^\text{cl}$&$7.875\times10^{-5}$\\

3 & $Q_\text{6, tf}^\text{cl}$&$2.110\times10^{-4}$&  $n_\text{tf}$ + $\kappa_\text{ld}$&$6.904\times10^{-5}$& $n_\text{ld}Q_\text{6, tf}^\text{cl}$&$7.970\times10^{-5}$\\
\bottomrule

\end{tabular} 
\label{tab:mle-op-1}
\end{table}

\begin{table}[h]
\caption{Top three ranked OPs for each choice of the trial RC ($\textbf{r}$) for the 12--6 system.}
\centering
    
\begin{tabular}{@{} l c c c c c c @{}} \toprule

& \multicolumn{2}{c}{$r_i=\alpha_1\times q_i+\alpha_0$} & \multicolumn{2}{c}{$r_{ij}=\alpha_2\times q_j +\alpha_1\times q_i+\alpha_0$} & \multicolumn{2}{c}{$r_{ij}=\alpha_1\times q_{ij}+\alpha_0$} \\
\cmidrule{1-3}
\cmidrule{4-5}
\cmidrule{6-7}
Rank & OP & -LL score  & OP & -LL score & OP & -LL score\\

\midrule
1 & $n_\text{tf}$&$1.180\times10^{-3}$&  $n_\text{tf}$ + $\kappa_\text{tf}$&$1.135\times10^{-3}$& $n_\text{tf}Q_\text{6, tf}^\text{cl}$&$1.130 \times 10^{-3}$\\

2 & $n_\text{ld}$&$1.228\times10^{-3}$&  $n_\text{tf}$ + $Q_\text{6, ld}^\text{cl}$&$1.460\times10^{-3}$& $n_\text{ld}Q_\text{6, ld}^\text{cl}$&$1.298\times10^{-3}$\\

3 & $Q_\text{6, ld}^\text{cl}$&$3.590\times10^{-3}$&  $n_\text{tf}$ + $f_\text{FCC}$&$1.490\times10^{-3}$& $n_\text{ld}Q_\text{6, tf}^\text{cl}$&$1.337\times10^{-3}$\\
\bottomrule

\end{tabular} 
\label{tab:mle-op-2}
\end{table}

From Table \ref{tab:mle-op-1} and \ref{tab:mle-op-2}, the linear combination of $n_\text{tf}$ and $Q_\text{6,tf}^\text{cl}$ is the trial RC that best correlates with the committor for both systems. However, models with a larger number of parameters will always achieve better likelihood.\cite{Peters_Obtaining_2006} As such, a Bayesian information criterion (BIC) was used to confirm if the addition of an OP (i.e., having two OPs instead of one) is a significant improvement in correlation with the committor. An addition of an OP is said to be a significant improvement to the model if: 

\begin{equation}
    LL_{2OP} - LL_{1OP} \geq BIC
\end{equation}
\begin{equation}
    BIC = \frac{1}{2}\ln(N_R) = 3.61
\end{equation}

where $LL_{2OP}$ is the log-likelihood score of two OPs model, $LL_{1OP}$ is the log-likelihood score of one OP model, and $N_R$ is the total number of configurations sampled. Equation 10 is not satisfied for both the 7--6 and 12--6 systems according to the data from Table \ref{tab:mle-op-1} and \ref{tab:mle-op-2}. As such, the linear combination of $n_\text{tf}$ and $Q_\text{6,tf}^\text{cl}$ is not a significantly better trial RC than the one OP model. Between the one OP model, the nonlinear combination of $n_\text{tf}$ and $Q_\text{6,tf}^\text{cl}$ provides the best log-likelihood score. Correspondingly, the MLE predicts this trial RC to correlate best with the committor. 

\subsection{MLE result with the inclusion of $f_{cloud}$}

While the nonlinear combination of $n_\text{tf}$ and $Q_\text{6,tf}^\text{cl}$ remains as the best RC for the 7--6 system, the nonlinear combination of $n_\text{ld}$ and $f_\text{clould}$ emerges as the best RC for the 12 --6 system. We choose the nonlinear form over the linear combination of two OPs (Model 2) for the same reasoning as provided above.

\begin{table}[h]
\caption{Top three ranked OPs for each choice of the trial RC ($\textbf{r}$) for the 7--6 system.}
\centering
      
\begin{tabular}{@{} l c c c c c c @{}} \toprule

& \multicolumn{2}{c}{$r_i=\alpha_1\times q_i+\alpha_0$} & \multicolumn{2}{c}{$r_{ij}=\alpha_2\times q_j +\alpha_1\times q_i+\alpha_0$} & \multicolumn{2}{c}{$r_{ij}=\alpha_1\times q_{ij}+\alpha_0$} \\
\cmidrule{1-3}
\cmidrule{4-5}
\cmidrule{6-7}
Rank & OP & -LL score  & OP & -LL score & OP & -LL score\\

\midrule
1 & $n_\text{tf}$&$5.760\times10^{6}$&  $n_\text{tf}$ + $f_\text{cloud}$&$5.236\times10^{6}$& $n_\text{tf}Q_\text{6,tf}^\text{cl}$&$5.686 \times 10^{6}$\\

2 & $n_\text{ld}$&$6.027\times10^{6}$&  $n_\text{tf}$ + $Q_\text{6,ld}^\text{cl}$&$5.534\times10^{6}$& $n_\text{ld}f_\text{cloud}$&$5.691\times10^{6}$\\

3 & $f_\text{cloud}$&$1.167\times10^{7}$&  $n_\text{tf}$ + $Q_\text{6,tf}^\text{cl}$&$5.620\times10^{6}$& $n_\text{ld}\kappa_\text{ld}$&$6.437\times10^{6}$\\
\bottomrule

\end{tabular} 
\label{tab:mle-op-3}
\end{table}

\begin{table}[h]
\caption{Top three ranked OPs for each choice of the trial RC ($\textbf{r}$) for the 12--6 system.}
\centering
    
\begin{tabular}{@{} l c c c c c c @{}} \toprule

& \multicolumn{2}{c}{$r_i=\alpha_1\times q_i+\alpha_0$} & \multicolumn{2}{c}{$r_{ij}=\alpha_2\times q_j +\alpha_1\times q_i+\alpha_0$} & \multicolumn{2}{c}{$r_{ij}=\alpha_1\times q_{ij}+\alpha_0$} \\
\cmidrule{1-3}
\cmidrule{4-5}
\cmidrule{6-7}
Rank & OP & -LL score  & OP & -LL score & OP & -LL score\\

\midrule
1 & $n_\text{tf}$&$5.787\times10^{6}$&  $n_\text{tf}$ + $f_\text{cloud}$&$5.133\times10^{6}$& $n_\text{ld}f_\text{cloud}$&$5.515 \times 10^{6}$\\

2 & $n_\text{ld}$&$6.027\times10^{6}$&  $n_\text{tf}$ + $\kappa_\text{tf}$&$5.570\times10^{6}$& $n_\text{tf}Q_\text{6,tf}^\text{cl}$&$6.034\times10^{6}$\\

3 & $f_\text{cloud}$&$1.167\times10^{7}$&  $n_\text{tf}$ + $Q_\text{6,ld}^\text{cl}$&$5.620\times10^{6}$& $n_\text{ld}Q_\text{6,ld}^\text{cl}$&$6.366\times10^{6}$\\
\bottomrule

\end{tabular} 
\label{tab:mle-op-4}
\end{table}
\newpage

\subsection{Histogram Test} \label{sec:hist-test}
\begin{figure}[h!]
    \centering
    \includegraphics[width=\linewidth]{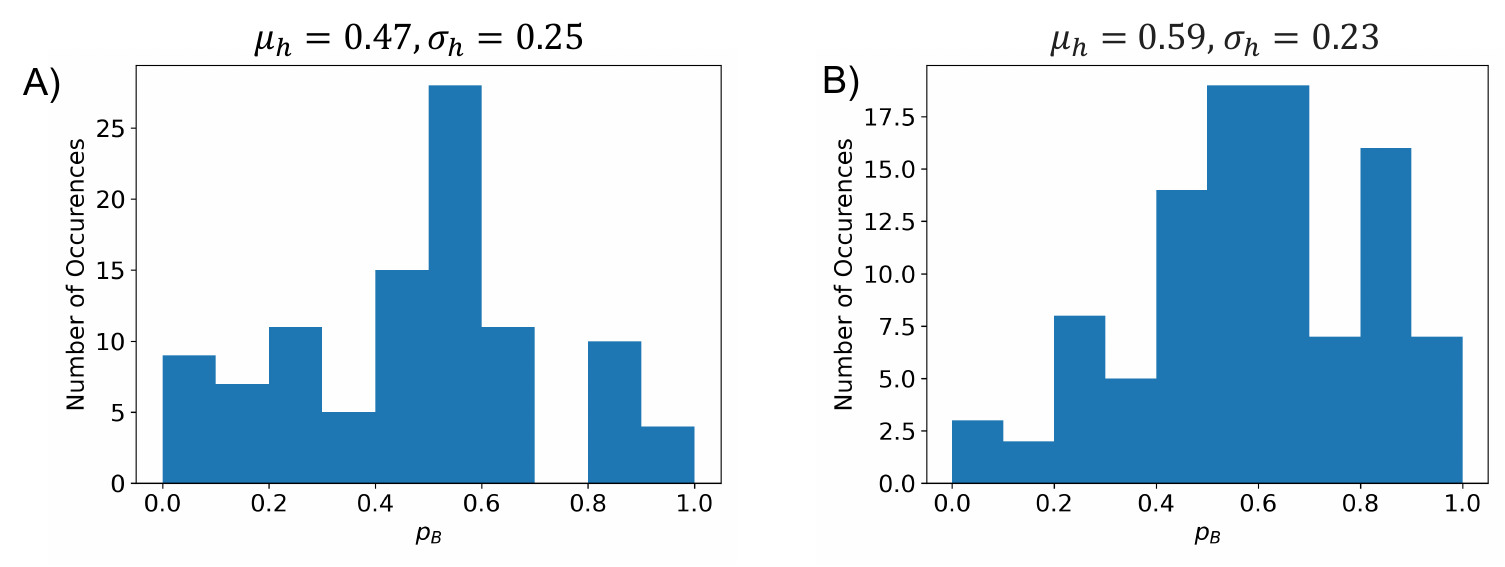}
    \caption{Histogram test for the transition state configurations for the (A) 7--6 system and (B) 12--6 system, where the trial RC is $n_{tf}Q_{6,tf}^{cl}$. The procedure for the test is identical for both systems. One hundred configurations with estimated $0.45 \leq p_B \leq 0.55$ from Fig. 3A\&B were randomly selected. For each configuration, we launched 20 momenta-randomized trajectories and observed whether the trajectory would melt back to liquid first (state A) or if it would keep growing until some solid size (state B). Then the true committor is calculated for each configuration as the fraction of trajectories that reached B before reaching A. $\mu_h$ and $\sigma_h$ signify the mean and standard deviation of the corresponding histogram.}
    \label{fig:hist-test}
\end{figure}

\begin{figure}[h!]
    \centering
    \includegraphics[width=\linewidth]{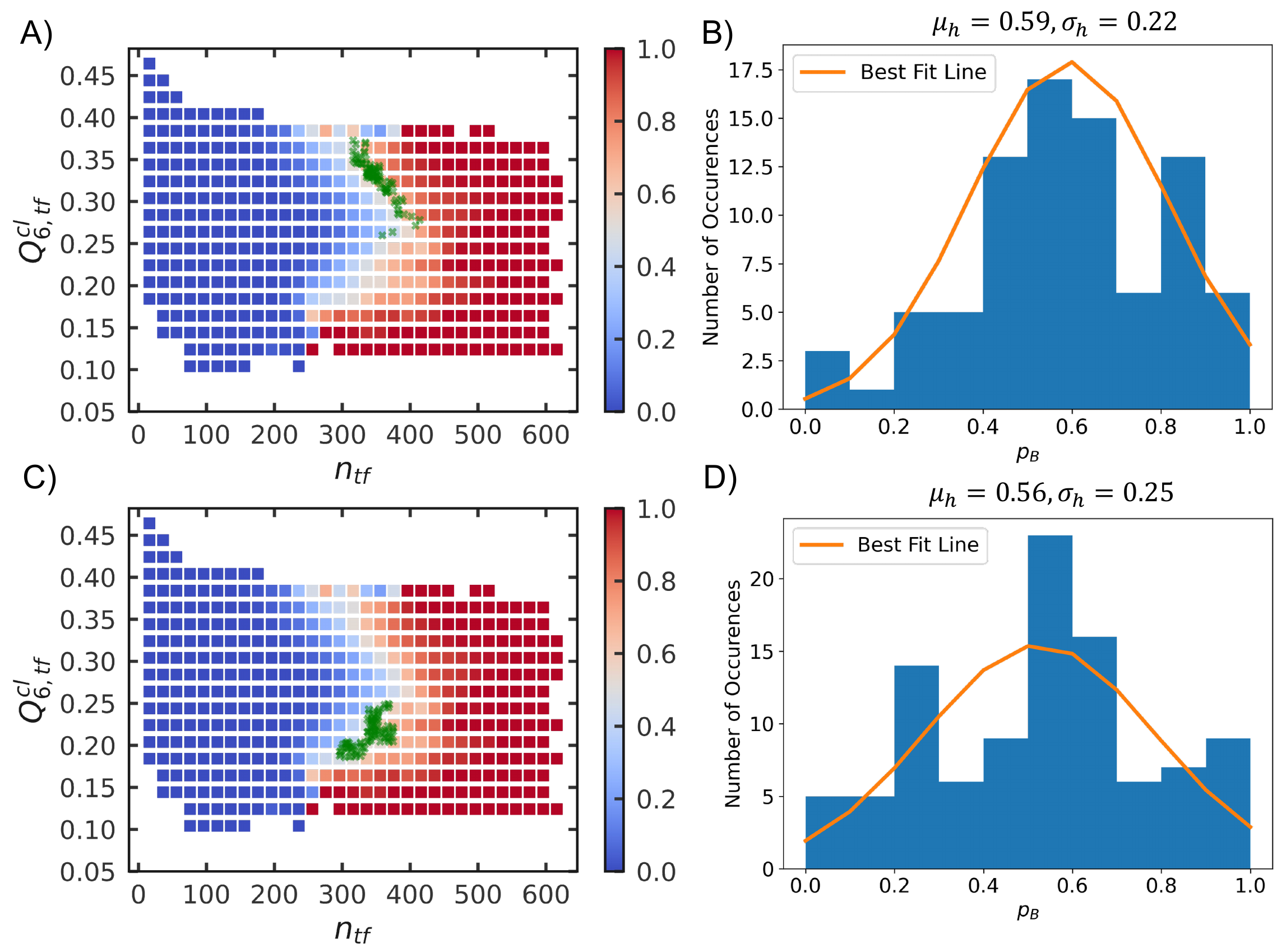}
    \caption{Histogram test for the configurations from the upper (A-B) and lower (C-D) elbow transition region for the 12--6 system. The green crosses on panels (A) and (C) signify the configurations that were used for the histogram tests (B) and (D), respectively.}
    \label{fig:up-low-hist-test}
\end{figure}

\begin{figure}[h!]
    \centering
    \includegraphics[width=\linewidth]{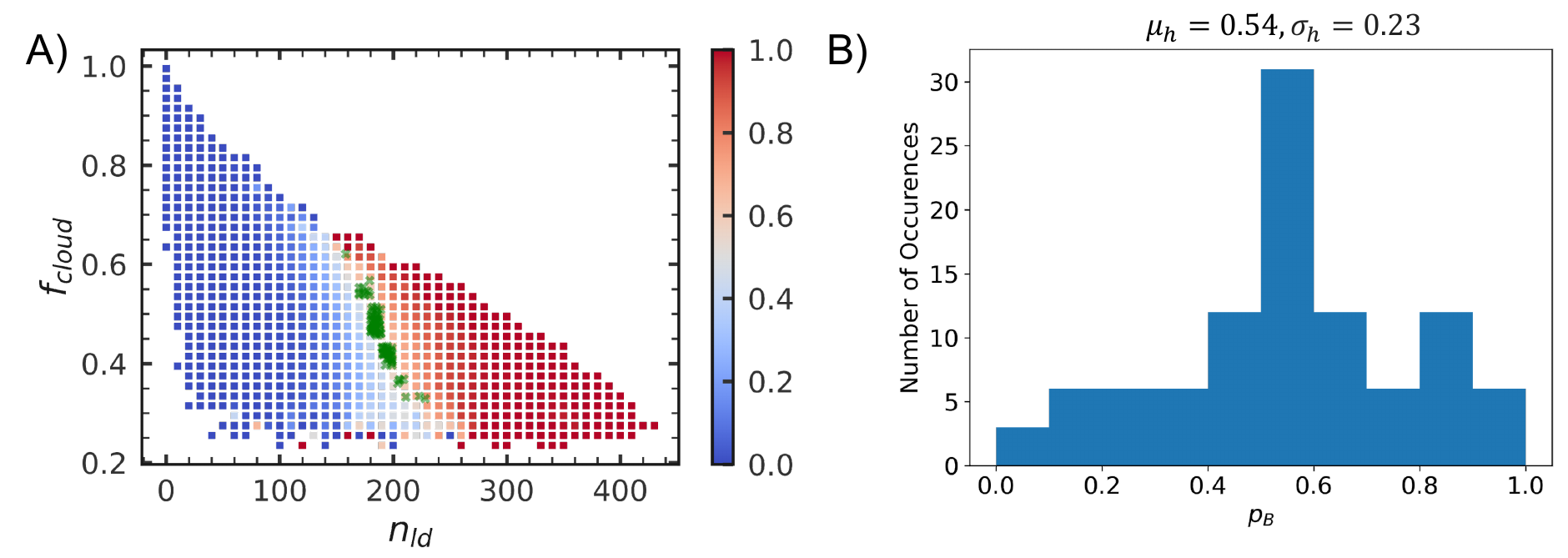}
    \caption{(A) Projection of $p_B$ onto $n_{ld}$ and $f_{cloud}$. The green crosses signify the configurations that were used for the histogram test. (B) Histogram test for the transition state configurations for the 12--6 system after f$_\text{cloud}$ was included in the MLE, where the trial RC is $n_{ld}f_{cloud}$. One hundred configurations with estimated $0.45 \leq p_B \leq 0.55$ from Fig. 4 were randomly selected. The rest of the procedure was the same as the figures above.}
    \label{fig:12-6_hist-test-new}
\end{figure}

We confirm the validity of the different trial RCs through the histogram test as described by Ref.~\citenum{Peters_Obtaining_2006} (see Fig.~\ref{fig:hist-test},~\ref{fig:up-low-hist-test} and \ref{fig:12-6_hist-test-new}). The procedure for the histogram test is the same for all trial RCs. First, we randomly select 100 configurations that are estimated to have $0.45 \leq p_B \leq 0.55$ according to each trial RC. Then, we randomize the configurations' velocities and extend 20 simulations from each configuration until the trajectories either reach $\lambda_A$ or $\lambda_B$. Afterward, we calculate the true committor probability of these configurations by determining the fraction of trajectories that reach $\lambda_B$ before $\lambda_A$. We then bin the configurations based on their committor probability and record the location of the distribution's peak. If the distribution peaks at $p_B\approx0.5$, the trial RC passes the histogram test and is used as true RC. 

\begin{figure}[h!]
    \centering
    \includegraphics[width=\linewidth]{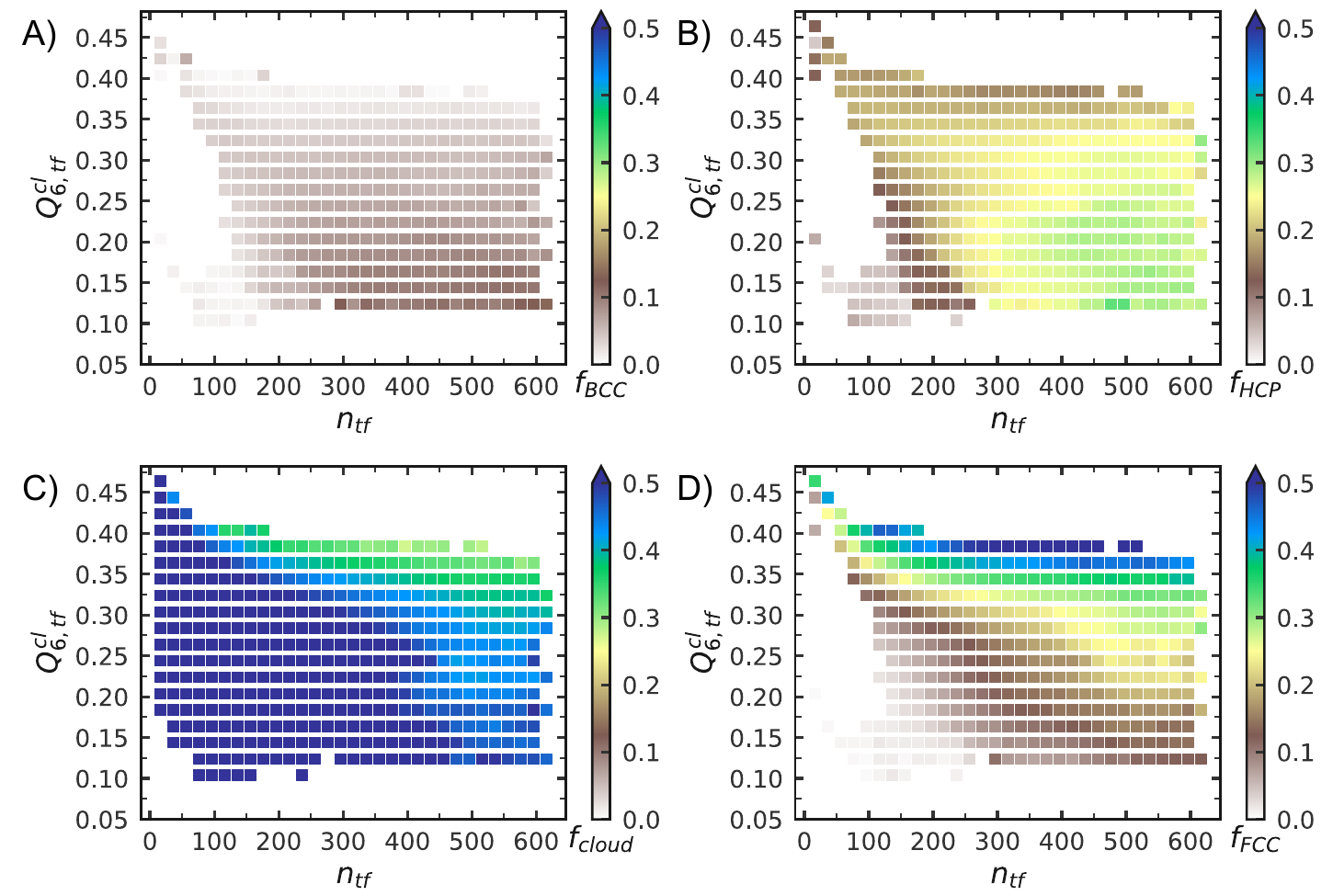}
    \caption{Projection of (A) $f_{BCC}$, (B) $f_{HCP}$, (C) $f_{cloud}$, and (D) $f_{FCC}$ onto $n_{tf}$ and $Q_{6,tf}^{cl}$ for the 12--6 system. The color bar is open-ended on the upper end to signal that there are values higher than 0.5 for some bins. Values higher than 0.5 are colored by dark navy blue. The color bar is plotted up to 0.5 to accentuate differences in visualization. Through panels (B) and (D), we observe that there is a correlation between the fraction of HCP/FCC particles and the lower/upper elbows of the transition region, respectively. This provides further explanation for the elbow we observe in Fig. 3B. This plot suggests that the existence of the lower elbow emerges from sampling nuclei with higher HCP content, which perhaps was not sampled in previous studies.\cite{Moroni_Interplay_2005, Beckham_Optimizing_2011}}
    \label{fig:126_comp_heatmap}
\end{figure}

\clearpage

\section{Variational Autoencoder (VAE)}
Details of the VAE architecture and features for the 7--6 system are described in Ref.~\citenum{SarupriaJCTC2025}. The VAE architecture used for the 12--6 is the same one as the 7--6. As for the local descriptors used in the 12--6, they are also similar to those of the 7--6, with exceptions to the r$_\text{cut}$ used. Here, the set of r$_\text{cut}$ values used were  1.3$\sigma$, 1.5$\sigma$, and 1.7$\sigma$.

\vspace{1cm}

\begin{figure}[h!]
    \centering
    \includegraphics[width=\linewidth]{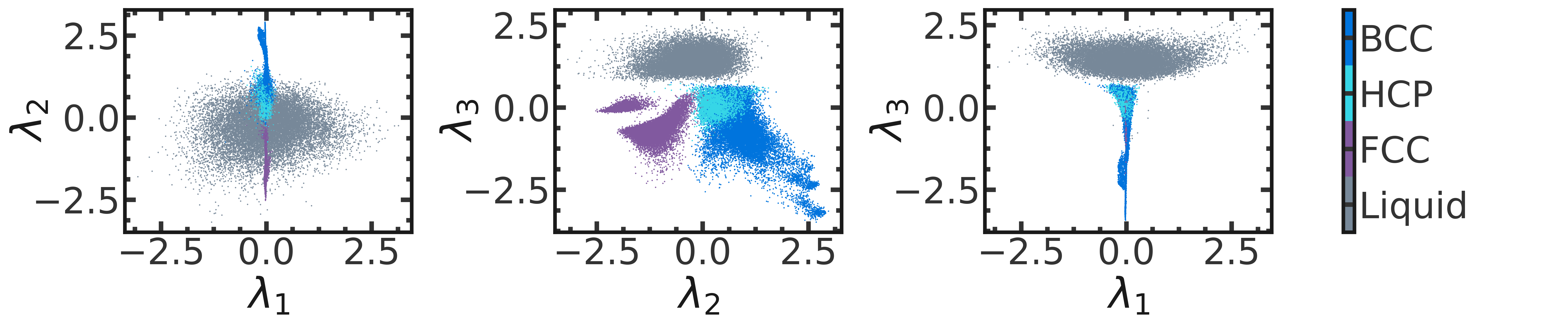}
    \caption{VAE projection of bulk solid particles from the 7--6 system. The overall latent space dimension is 3. Each panel represents a 2D projection of the particles in the latent space. The label of each particle is acquired based on the values of $\overline{q}_{4}$ and $\overline{q}_{6}$. Blue points correspond to BCC, cyan points to HCP, and purple points to FCC.}
    \label{fig:76vae}
\end{figure}

\begin{figure}[h!]
    \centering
    \includegraphics[width=\linewidth]{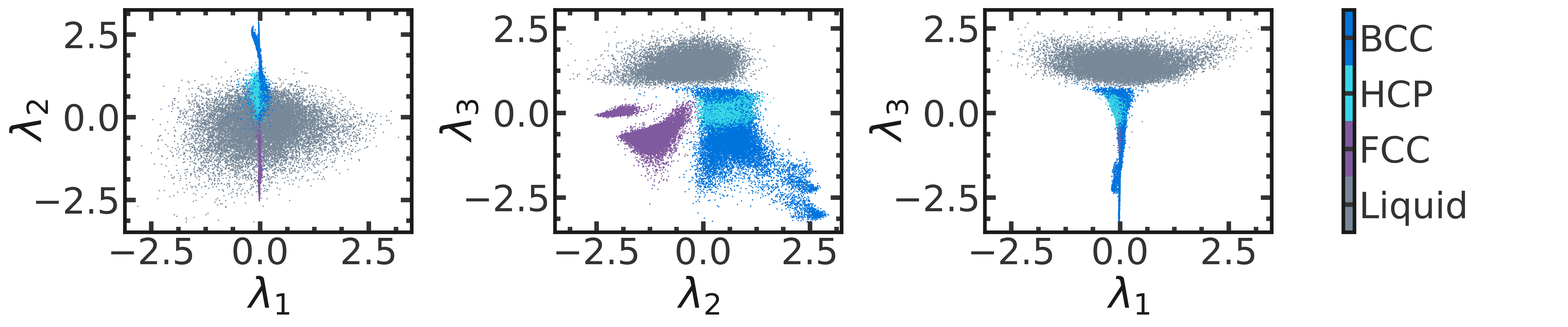}
    \caption{VAE projection of bulk solid particles from the 12--6 system.}
    \label{fig:126vae}
\end{figure}

\clearpage


\begin{figure}[h!]
    \centering
    \includegraphics[width=\linewidth]{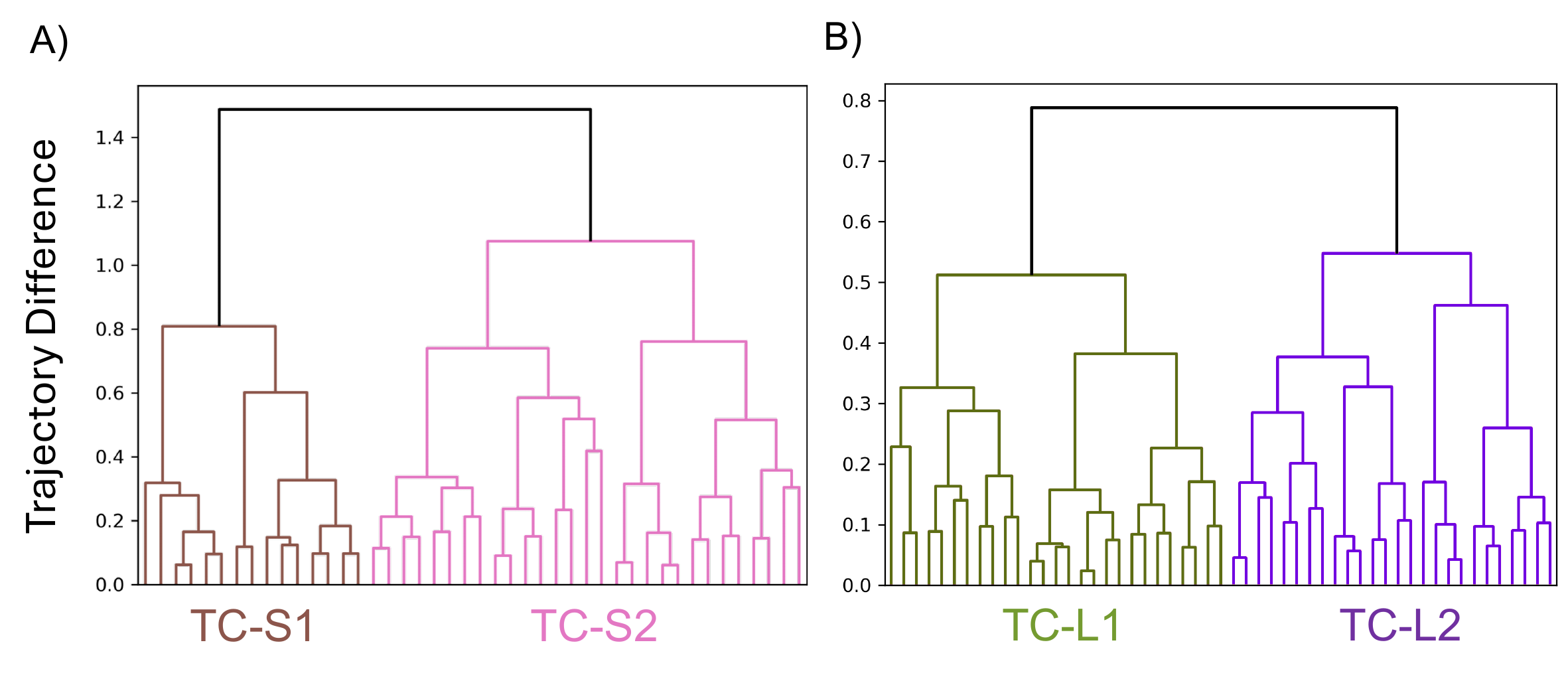}
    \caption{Hierarchical agglomerative clustering of (A) 7--6 and (B) 12--6 nucleation trajectories. Each colored region of each dendrogram represents a cluster of nucleation trajectories. The 7--6 trajectory clusters consist of TC-S1 and TC-S2. The 12--6 trajectory clusters consist of TC-L1 and TC-L2. Nucleation trajectories that share the most similarity will join branches earlier as the cutoff (y-axis) is increased.}
    \label{fig:nuccls}
\end{figure}

\begin{figure}[h!]
\centering
\includegraphics[width=\linewidth]{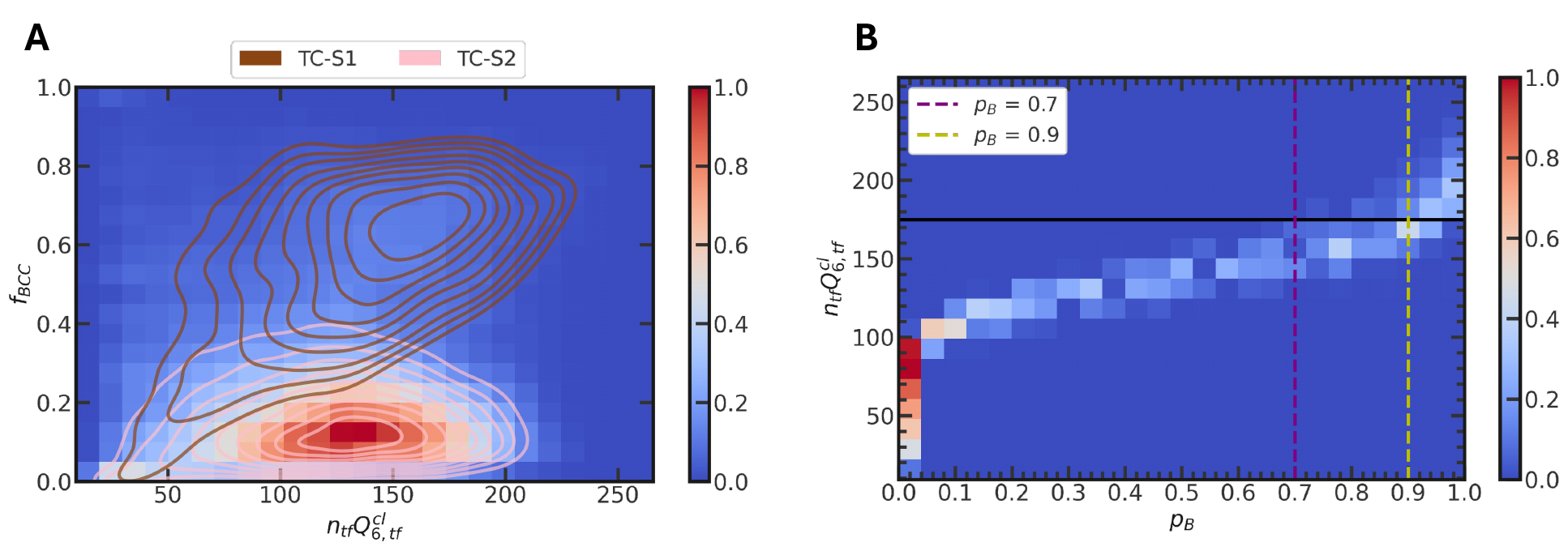}
\caption{(A) Normalized density projection of the 7-6 nucleation trajectories onto the fraction of BCC in the nucleus ($f_{BCC}$) and the reaction coordinate ($n_{tf}Q_{6,tf}^{cl}$). The contour plots, labeled ``TC-S1" and ``TC-S2", highlight the density of trajectories belonging to the TC-S1 and TC-S2 trajectory clusters, respectively. (B) Normalized density projection of the 7-6 nucleation trajectories onto the reaction coordinate ($n_{tf}Q_{6,tf}^{cl}$) and the committor probability ($p_B$). The black horizontal line is the estimate for the lowest $n_{tf}Q_{6,tf}^{cl}$ value where the two trajectory clusters separate. The dashed vertical lines represent where $p_B$ is 0.7 (purple) and 0.9 (yellow). The color bar in each plot represents the normalized density of configurations along coordinates of interest. A higher value represents higher density, and a lower value represents lower density. }
\label{fig:si-stopframe}
\end{figure}

\begin{figure}[h!]
    \centering
    \includegraphics[width=\textwidth]{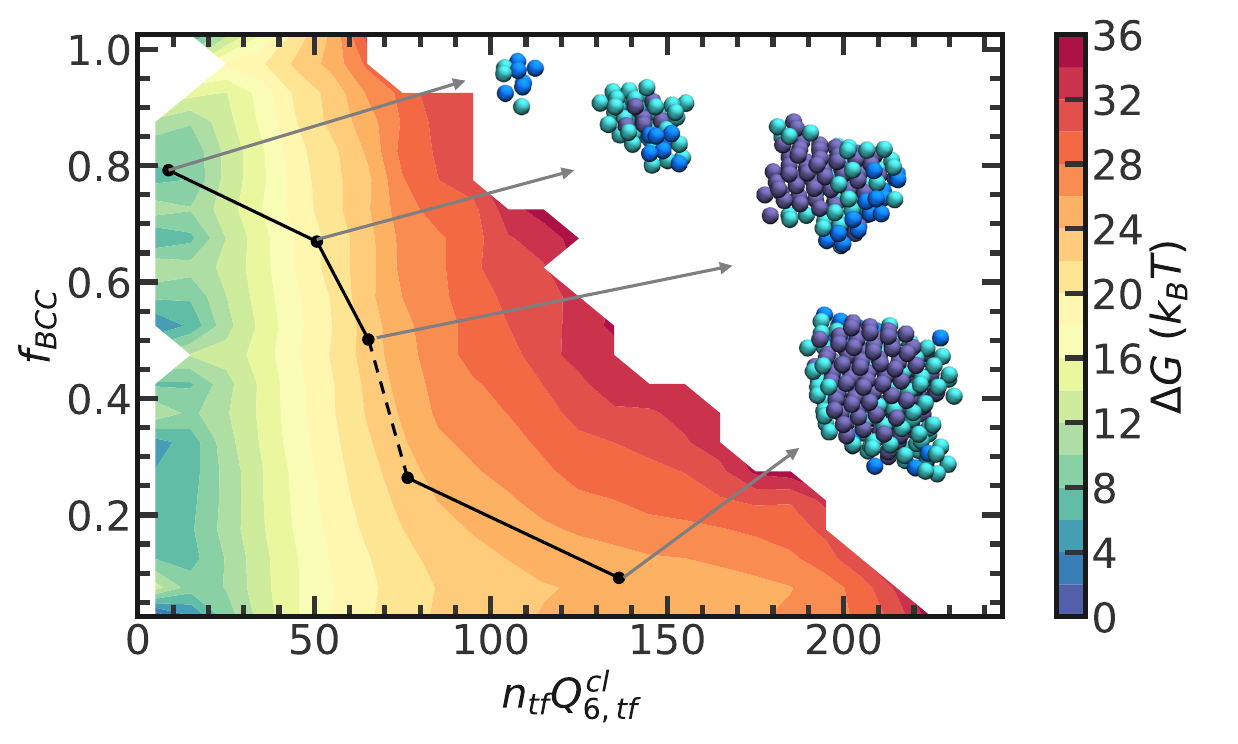}
    \caption{Conditional free energy of the 12--6 system along the RC and $f_\text{BCC}$.}
    \label{fig:126-condfe}
\end{figure}

\clearpage
\begin{movie}[MD trajectory of nucleation for the 7--6 system (sample 1).] The particles in the largest solid nucleus in the movie are colored based on their $\overline{q}_{4}-\overline{q}_{6}$ values. This trajectory shows a nucleation trajectory from TC-S1, which demonstrates the competition between HCP and BCC, where BCC eventually captures the nucleus core and dominates the composition.
\end{movie}

\begin{movie}
    [MD trajectory of nucleation for the 7--6 system (sample 2).] The particles in the largest solid nucleus in the movie are colored based on their $\overline{q}_{4}-\overline{q}_{6}$ values. This trajectory shows a nucleation trajectory from TC-S2, which demonstrates the competition between HCP and BCC, where HCP eventually captures the nucleus core and FCC cross-nucleates from HCP to form an FCC-HCP dominant composition.
\end{movie}

\clearpage

\providecommand{\latin}[1]{#1}
\makeatletter
\providecommand{\doi}
  {\begingroup\let\do\@makeother\dospecials
  \catcode`\{=1 \catcode`\}=2 \doi@aux}
\providecommand{\doi@aux}[1]{\endgroup\texttt{#1}}
\makeatother
\providecommand*\mcitethebibliography{\thebibliography}
\csname @ifundefined\endcsname{endmcitethebibliography}
  {\let\endmcitethebibliography\endthebibliography}{}